\documentclass[global,twocolumn,final,numbook]{svjour}
\usepackage{graphicx}
\usepackage[pdftex, colorlinks=true, pdfstartview=FitH,linkcolor=blue, citecolor=blue, urlcolor=blue]{hyperref}
\usepackage{color}
\usepackage[ruled,section]{algorithm}
\usepackage{algorithmic}
\usepackage{latexsym}
\usepackage{amssymb,amsfonts,amsbsy,amsmath,amscd}
\usepackage{float}
\usepackage[FIGBOTCAP,normal,bf,tight]{subfigure}
\usepackage{stmaryrd}
\usepackage{enumerate}
\usepackage{upgreek}
\usepackage{xcolor}
\newcommand{\VTK}[1]{\textsf{VTK}#1}
\newcommand{\PV}[1]{\textsf{ParaView}#1}
\newcommand{\Love}[1]{\textsf{Love}#1}

\graphicspath{{./pdf/}{./png/}{./plots/}}
\begin{document}
\title{Visualization and Analysis of Large-Scale, Tree-Based, Adaptive Mesh
Refinement Simulations with Arbitrary Rectilinear Geometry}
\author{Gu\'enol\'e \textsc{Harel}\inst{1}
	\and
        Jacques-Bernard \textsc{Lekien}\inst{1}
	\and
	Philippe P. \textsc{P\'eba\"y}\inst{2}
}
\institute{
        CEA, DAM, DIF, F-91297 Arpajon, France\\
        \email{\{guenole.harel,jacques-bernard.lekien\}@cea.fr}
	\and
	Positiveyes, 84330 Le Barroux, France\\
	\email{philippe.pebay@positiveyes.fr}
\date{} 
}
\def\makeheadbox{} 
\maketitle
\begin{abstract}
We present here the first systematic treatment of the
problems posed by the visualization and analysis of large-scale,
parallel adaptive mesh refinement (AMR) simulations on an Eulerian
grid.

When compared to those obtained by constructing an intermediate
unstructured mesh with fully described connectivity, our primary
results indicate a gain of at least 80\% in terms of memory footprint,
with a better rendering while retaining similar execution speed.

In this article, we describe the key concepts that allow us to obtain
these results, together with the methodology that facilitates the
design, implementation, and optimization of algorithms operating
directly on such refined meshes.
This native support for AMR meshes has been contributed to the open
source Visualization Toolkit (VTK).

This work pertains to a broader long-term vision, with the dual goal
to both improve interactivity when exploring such data sets in 2 and 3
dimensions, and optimize resource utilization.
\end{abstract}
\tableofcontents
\section{Introduction}
\label{s:introduction}
\subsection{Preamble}
Massive numerical simulations are nowadays
routinely run on petascale supercomputers such as
Tera100 and the pre-exascale Tera1000~\cite{tera:CEA}.
Among simulation codes, those using adaptive mesh refinement (AMR) are
especially efficient at tracking fine details within very large
domains of interest.
AMR enables a trade-off between numerical accuracy, memory footprint,
and computational cost, by allowing for mesh refinement (and
coarsening) in sub-regions of the simulation.
Recent large scale simulations have reached ten trillion
cells on a regular Eulerian grid~\cite{woodward:12}. 
This pioneering work, representative of what
will be ``everyday'' tomorrow exascale computing, showed that
it might however be either impossible to store due to a too large
number of elements, or would be computationally too expensive to
post-process.

\begin{figure}[htb]
\centering
\includegraphics[width=0.9\columnwidth]{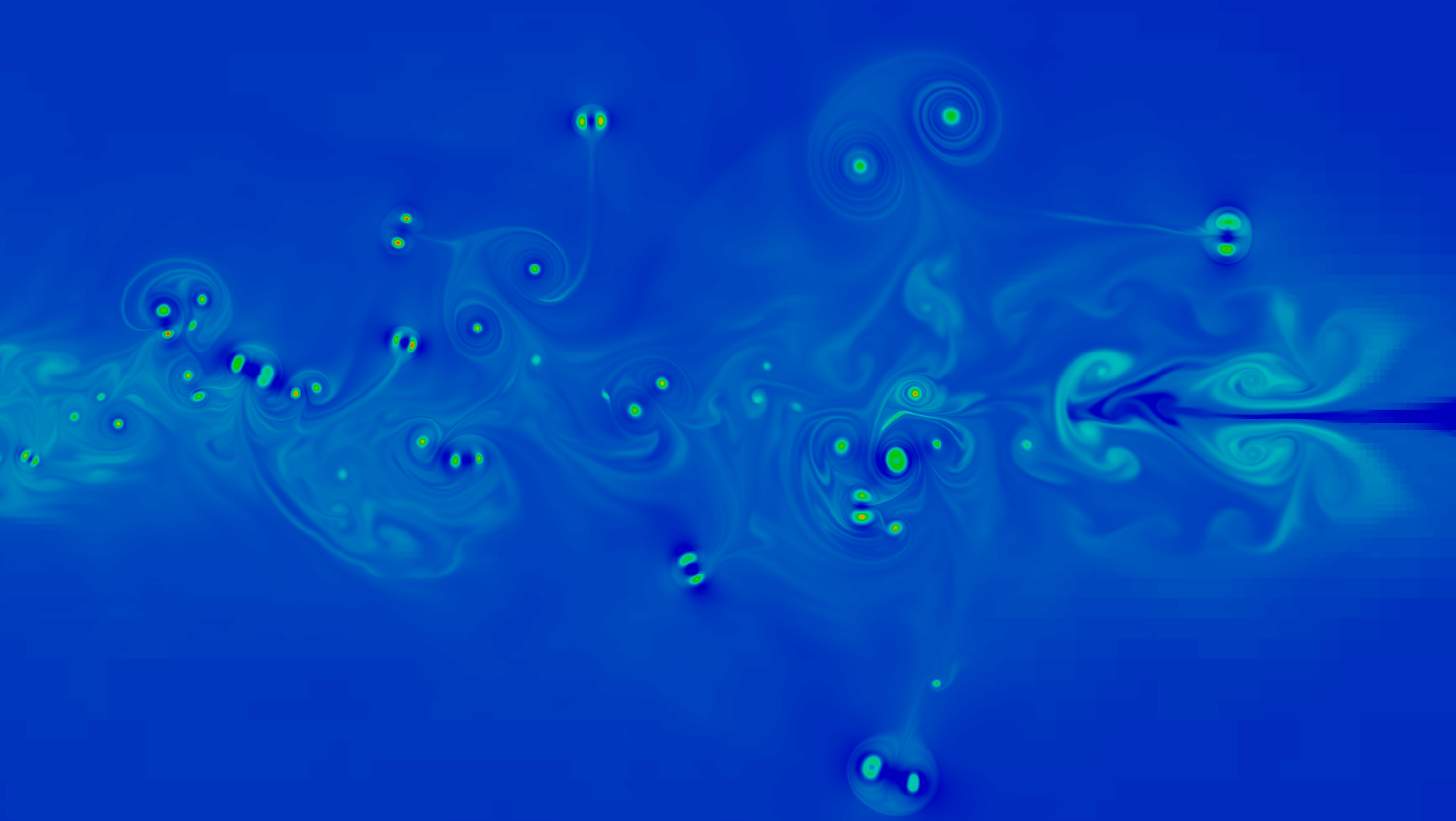}
\caption{Visualization of supersonic shockwave drag, simulated
on a tree-based AMR mesh.}
\label{fig:intro}
\end{figure}
These difficulties can be alleviated by refining the
original mesh only where needed, while retaining coarser elements
wherever local feature scales permit.
Of course, this approach is limited to those specific physical
problems where the meaningful phenomena are spatially localized.
This is the case, for example, in astrophysics~\cite{ramses:02},
transient wave propagation~\cite{lomov:05}, or shock wave
computation~\cite{bale:02,berger:11}, illustrated in
Figure~\ref{fig:intro}, all cases where some appropriate coarsening
criterion can also be easily defined.

Since the first description of an AMR methodology with the
Berger-Oliger~\cite{berger:84} type, several implementations have been
proposed and developed.
It is beyond the scope of this article to provide an in-depth
comparison of \emph{block structured} (also known as
\emph{patch-based}) versus \emph{tree-based} (also known as
\emph{point-wise structured}) AMR methodologies.
Nonetheless, in order to fully understand the motivations and
constraints of the work presented hereafter, one must be aware that
the fundamental difference between the two approaches is, essentially,
a trade-off between memory footprint, and complexity of processing
algorithms.
Specifically, \emph{ceteris paribus}, a typical tree-based AMR grid 
will occupy much less memory estate than its block structured equivalent,
at the cost of higher processing time as a result of more complicated
algorithms.
In fact, this dichotomy between methods arises from the more general
opposition between implicit and explicit representations, with the
ensuing consequences when storing, as opposed to processing, the
resultant data objects.
It is worth noticing here that the FLASH framework~\cite{flash:00}
offers both options, although this is actually done with two different
underlying codes: Chombo (block structured) and Paramesh (tree-based).

Block structured AMR will not be discussed in the rest of this
article; the interested reader can refer in particular to the Chombo
pages for more details~\cite{chombo:overview}.
Our interest instead focused on the analysis of data sets produced by
tree-based AMR codes.
Several codes pertain to this group, for instance starting with
successive refinement in octants (therefore producing \emph{octrees})
of an initial root cell as done in RAMSES~\cite{ramses:02}, or using a uniform,
structured grid of root cells as done in RAGE/SAGE~\cite{gittings:08} or
HERA~\cite{jourdren:05}, the tree-based AMR hydrodynamics simulation
code developed at CEA.

\subsection{Scope}
\label{s:scope}
We begin, in~\S\ref{s:context}, by providing the background and
context for this work, analyzing the challenges posed to scientific
visualization by tree-based AMR simulations.
As a result, we propose our global vision for addressing these
challenges in an exascale perspective.
However, the scope of this article is limited to the foundational
aspects of this vision, by means of laying out the necessary data
structures as well as the methodology to optimally process these.

What the reader will get from reading~\S\ref{s:foundations} is a
full understanding of our novel data structures, which we implemented
in \VTK{}~\cite{avila:10}.
Wherever necessary, based in particular on acquired experience with
large-scale data sets, we mention changes to claims or hypotheses
which we had made in earlier work~\cite{carrard:imr21}.

We then study in~\S\ref{s:method} the method we
designed to operate on these data objects, with a particular emphasis
on execution speed, in order to maintain interactivity even with the
largest data sets that can be stored on currently available
hardware.

We illustrate this methodology in~\S\ref{s:isocontouring}, with a case
of particular interest, namely iso-contouring, which is arguably
one of the most widely used visualization techniques.
It is also the most complicated, amongst all algorithms we have
devised and implemented so far for our tree-based data sets, due
to the intrinsic complications that arise from the very topological
nature of iso-manifolds. 

In~\S\ref{s:results}, we examine the validity of our claims relative
to performance with a set of tests, that are representative of the
scientific simulation data sets we wish to address.

Finally, we conclude this article by examining to what extent the
work presented in these pages covers what we initially intended to do.
We subsequently discuss how future work will be articulated with what
has been achieved so far, in order to achieve our long-term vision.

\section{Context}
\label{s:context}
\subsection{Problem Statement}
\label{s:problem}
In order to exploit the massive data sets produced by the
various numerical simulation codes of CEA, our visualization team
developed the Large Object Visualization Environment
(\Love)~\cite{aguilera:07,love:12}, a dedicated parallel visualization
tool.
It based on \VTK/\PV, an open-source, \textsf{C++} set of libraries an
applications for scientific data visualization and analysis supporting
many data types and featuring hundreds of algorithms, with thousands
of users in the global scientific community.

One approach for the visualization and analysis of AMR data sets with
\VTK{} is to use its native unstructured grid data objects.
One obvious advantage of this method is to make available the wealth
of existing filters already available in \VTK{} for such data sets
(e.g., cutting, clipping, iso-contouring, etc.).
However, the additional memory requirements that arise from converting
a mostly implicit data object into a fully explicit one rapidly
become prohibitive as the size of the grid grows.
Furthermore, when the cells of an AMR mesh are directly used as
unstructured element inputs (quadrilaterals or hexahedra) of an
algorithm such as iso-contouring, topological irregularities resulting
in strong visual artifacts such as gaps may appear.
These are caused by the topology of AMR meshes, which have
partly connected vertices (``T-junctions'') when they
contain neighboring cells at different refinement levels.
Linear interpolation, commonly used by visualization algorithms,
produces discontinuities across T-junctions, ultimately resulting in
incorrect visualizations.

\begin{figure}[htb]
\centering
\begin{minipage}[t]{0.45\columnwidth}
\centering
\vspace{0pt}
\includegraphics[width=0.9\columnwidth]{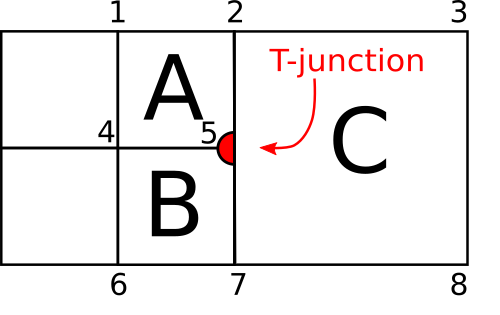}\\[2mm]
\includegraphics[width=0.9\columnwidth]{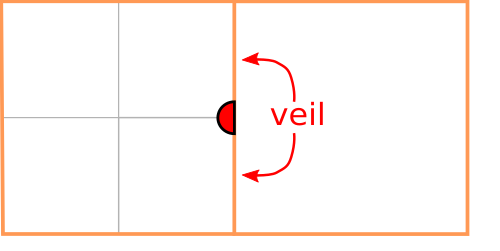}\\[2mm]
\includegraphics[width=0.9\columnwidth]{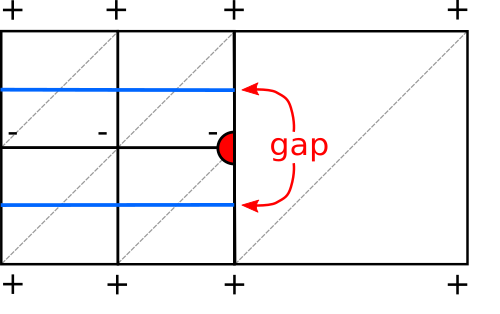}
\end{minipage}
\hfil
\begin{minipage}[t]{0.45\columnwidth}
\centering
\vspace{0pt}
\includegraphics[width=0.9\columnwidth]{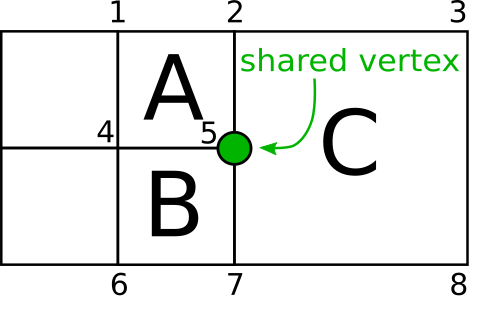}\\[2mm]
\includegraphics[width=0.9\columnwidth]{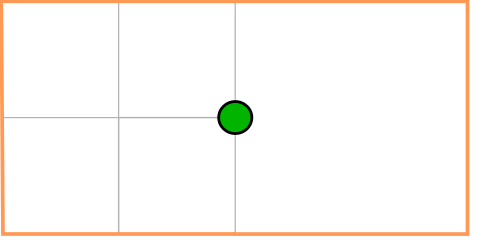}\\[2mm]
\includegraphics[width=0.9\columnwidth]{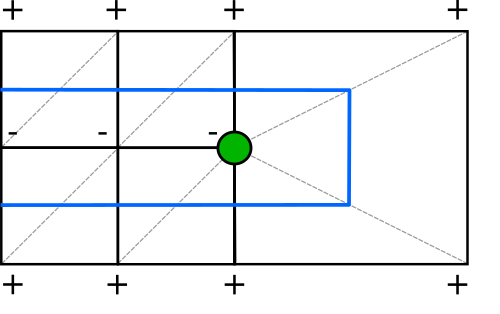}
\end{minipage}
\caption{Top: AMR grid converted into: (left) a quadrangle mesh with a
T-junction at point 5, which is shared by A and B but not by C, and
(right) a generic unstructured mesh where vertex 5 is shared by
pentagon C as well as  quadrangles A and B.
Middle: outside boundary (orange) computed over these 2 meshes; a topological
artifact (\emph{veil}) is caused by the T-junction (left), but not in
the conforming, generic mesh (right).
Bottom: linear iso-contour (blue) computed over theses 2 meshes,
between vertex values above ($+$) or below ($-$) a given value; dashed
gray lines represent possible triangulations used by the contouring
algorithm.} 
\label{fig:AMR-Unstructured-2D}
\end{figure}
The latter problem is further explicated in dimension~2 by
Figure~\ref{fig:AMR-Unstructured-2D}, where the top row represents
an AMR grid with 5 cells.
On the left, the cells are considered as the elements of an
unstructured quadrilateral mesh: by construction, quadrangle C does not
have any reference to vertex 5, creating a T-junction along edge
2--7.
On the right, the cells are now viewed as arbitrary polygons, with 
pentagon C sharing vertex 5 with quadrilaterals A and B, hence
eliminating the T-junction. 
Attempting to extract the outside boundary of the quadrilateral mesh
results in a topological artifact, called a \emph{veil}, whereas the
outside boundary is correctly extracted on the polygonal mesh.
Similarly, the effect of linear iso-contouring on both constructions,
when cell values are above or below a given iso-value are shown in the
bottom row.
The T-junction on the left causes a gap in the iso-contour, because
the algorithm cannot detect a contour intercept along edge 2--7 of
cell C.
Meanwhile, the same contouring algorithm is able correctly process the
generic cells, and produces a correct iso-contour without false gaps.

\begin{figure}[htb]
\centering
\begin{minipage}[t]{0.45\columnwidth}
\centering
\vspace{0pt}
\includegraphics[width=0.99\columnwidth]{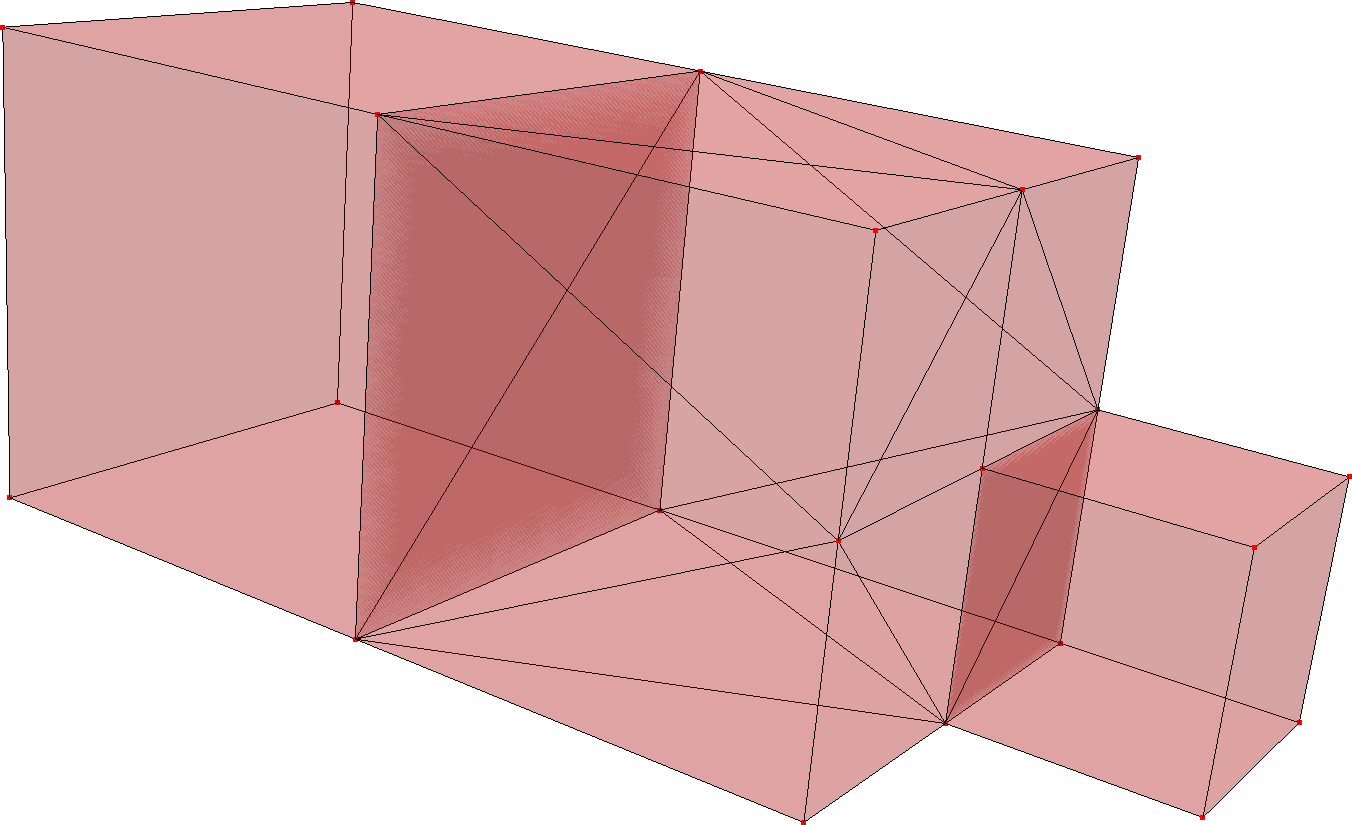}\\[2mm]
\includegraphics[width=0.99\columnwidth]{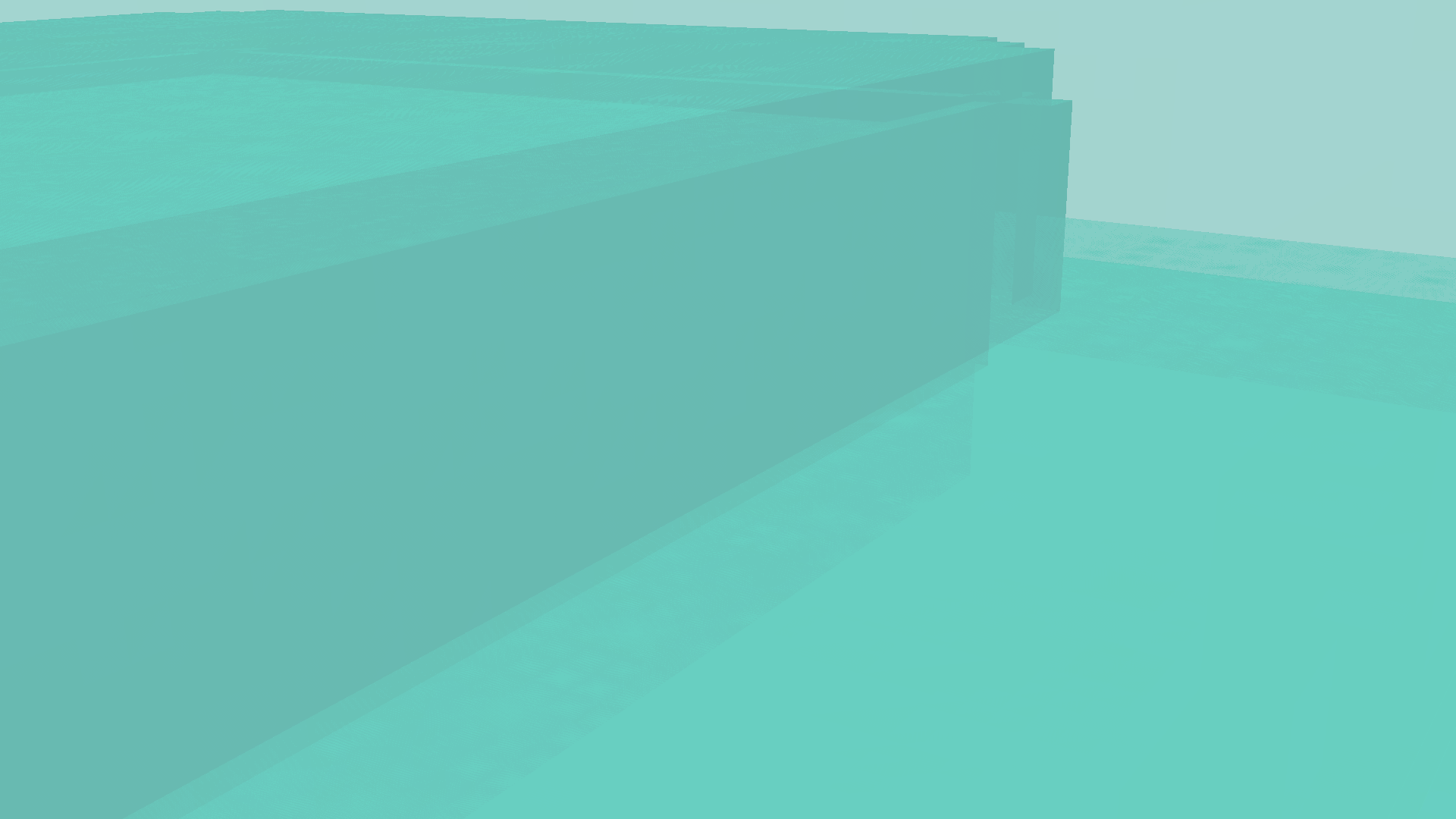}
\end{minipage}
\hfil
\begin{minipage}[t]{0.45\columnwidth}
\centering
\vspace{0pt}
\includegraphics[width=0.99\columnwidth]{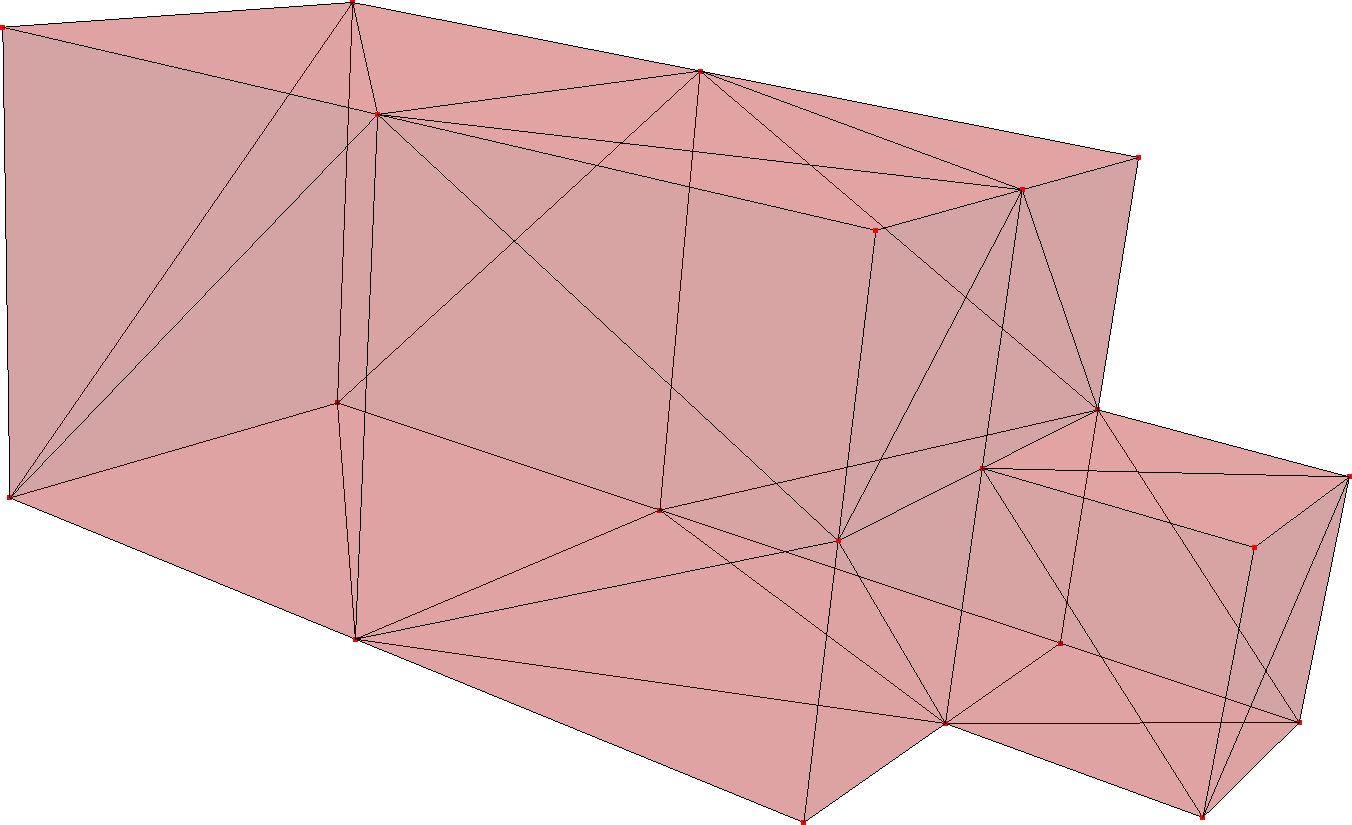}\\[2mm]
\includegraphics[width=0.99\columnwidth]{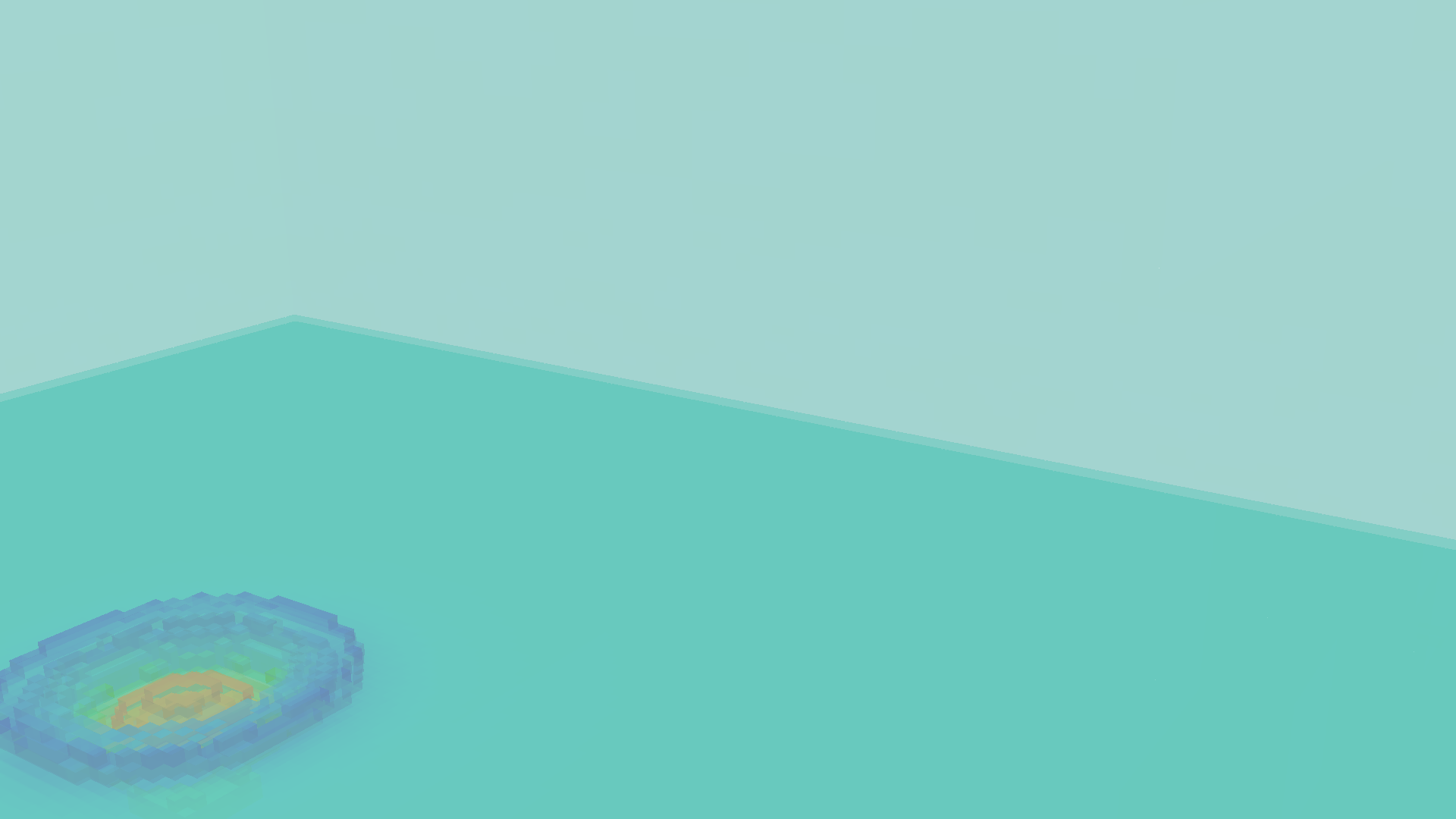}
\end{minipage}
\caption{Top row, left: one generic cell located between 2 hexahedra,
resulting in the appearance of extraneous veils being generated by the
\VTK{} geometry filter; right: all cells are generic and the boundary
is correctly extracted by the filter.
Bottom row: outside boundary of a bi-material 3D AMR simulation; left:
extraneous veils appear when the filter is applied to a mixed-cell
conforming unstructured mesh; right, the boundary is correctly
extracted with all generic cells.}
\label{fig:AMR-Unstructured-3D}
\end{figure}
The Hercule I/O library developed at CEA~\cite{hercule:12} supports
such conversion from AMR grids into unstructured, conforming meshes.
Prior to 2012, this was the only option available to visualize the
tree-based AMR data sets produced at CEA.
In addition to the already discussed performance limitations, using
this approach also comes at the price of reduced interactivity because
of I/O latency.

Furthermore, \VTK{} does not support well the mixing of hexhaedral
elements with generic cells as illustrated in
Figure~\ref{fig:AMR-Unstructured-3D}, left.
Specifically, when attempting to extract the outside surface of the
unstructured mesh with mixed cells, the subdivision of the generic
cells by \VTK{} results in incompatible tessellations across
neighboring element faces.
Although it is possible to resolve this problem by using only generic
cells, as shown in Figure~\ref{fig:AMR-Unstructured-3D}, right, but
the computational and memory costs quickly become prohibitive for
realistically-sized meshes.

\begin{figure}[htb]
\centering
\includegraphics[width=0.95\columnwidth]{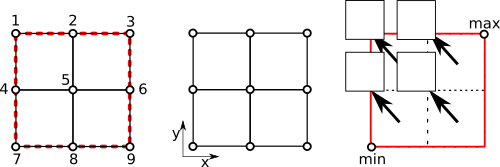}
\caption{Two different representations of the same mesh: explicit
unstructured representation (left), versus AMR description (right).
The red color coding indicates the root cell, which is stored in the
AMR representation, but not in the unstructured mesh.}
\label{fig:UnstructuredVsAMR2D}
\end{figure}
It is easy to illustrate, for instance with the simple example
depicted in Figure~\ref{fig:UnstructuredVsAMR2D}, the dramatic
inefficiency of using explicit unstructured meshes to represent AMR
grids.
Considering a quadrilateral in dimension 2, decomposed into 4
sub-elements, it is straightforward to devise a corresponding
tree-based AMR representation using $4$ floats for the extremal
coordinates of the grid and $1$ Boolean value to indicate that the
quadrangle is subdivided.
Meanwhile, an explicit unstructured representation of the
same requires $9\times2=18$ floats for the vertex
coordinates, as well as $4\times4=16$ integers to describe the
connectivity of the $4$ cells.
Therefore, the AMR description reduces the memory footprint of almost
a full order of magnitude for this simple case alone.

\VTK{} has long provided some support for block-structured AMR data
sets.
Prior to 2012, it also offered very limited support for a
particular case of tree-based AMR with a single-root octree
object~\cite{yau:83}. 

\begin{figure}[htb]
\centering
\includegraphics[width=0.95\columnwidth]{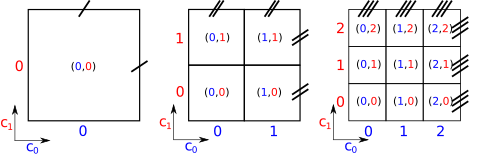}
\caption{The 3 allowed AMR subdivision patterns in dimension~$2$: without
refinement ($f=1$, left), binary subdivision ($f=2$, center), and ternary
subdivision ($f=3$, right), with respective numbers of children equal
to $1$, $4$, and $9$. In dimension~3 these translate respectively into
$1$, $8$, and $27$ children.}
\label{fig:BranchingFactors2D}
\end{figure}

Furthermore, simulation codes such as HERA use either binary or
ternary subdivision schemes when refining meshes, i.e.,
with \emph{branching factor} $f\in\{2;3\}$ along each dimension of the
grid.
This is illustrated by Figure~\ref{fig:BranchingFactors2D} in
dimension~$2$.
In general, in dimension $d$, refining a cell results
in obtaining $f^d$ sub\emph{children} (sub-cells).
Any post-processing methodology designed to handle the results of such
simulations must therefore be able to accommodate, not only the usual
binary trees, quadtrees and octrees, but also more exotic ternary
trees.
Finally, another constraint to be taken account is the fact that AMR
simulation codes used at CEA are run in parallel, with the
corresponding data sets being distributed over many thousands of
compute nodes.
These codes balance computational sub-domains by allocating the root
cells in the grid of trees, resulting in individual AMR trees that
are never shared between different compute nodes.
Traversal objects for such grids of trees must therefore be carefully
designed in order to \emph{a priori} allow for extremely unbalanced
trees structures between various areas of the overall domain.

\subsection{Vision}
\label{s:vision}
Our global, long-term vision for tree-based AMR visualization and
analysis can be articulated as follows:
\begin{enumerate}[\bf{[}a{]}]
\item
Propose a novel \VTK{} data object to support all requested tree-based
features, that is both memory-efficient and able to convert such objects
into conforming meshes. This is to allow for the direct
utilization of the wealth of existing unstructured mesh algorithms
when explicitly requested.
\item
Design and implement visualization and analysis algorithms that are
specific to the primary tree structure, as needed by actual users,
with a strong emphasis on performance.
In our vision, this optimization of execution speed is best achieved
by using specialized constructs called \emph{cursors} and
\emph{supercursors}.
\item
Optimize rendering speed, in order be able to maintain interactivity
when visualizing the largest possible tree grids that can be contained
in memory. 
A possible approach could be to take advantage of the tree structure
of the grids, to allow for level-of-detail culling relative to the
size of the rendering window, screen resolution, view and camera
position, etc.
\item
Qualitatively improve the final rendering with, e.g., mapping, texture
splatting or ray tracing techniques specifically tailored for the
tree-based AMR objects.
\item
Design and implement a way to pass object information, so a reader
specific to tree-based AMR grids  be able to limit actual reading and
storing of those parts of the entire grid that are explicitly needed
by filters and rendering (such as maximum depth of refinement and
bounding box).
\item
Define a serialization specification for these structures, and develop
I/O classes implementing it.
Such a serialization protocol will also improve current parallel load
balancing schemes by allowing for communication of
large sub-grids in small messages.
\item
Expand the range of supported tree-based AMR data sets; envisioned
objects include grids that have many root cells but a small number of
refinement levels or, conversely, that only have a very small number
of root cells with many refinement levels.
Such extensions would have to be achieved while maintaining the same
goal of memory footprint minimization and execution speed maximization.
\item
Expand the current post-processing paradigm to include concurrent
approaches based on \emph{in situ} and \emph{in transit}
processing.
Such a data-centric approach would allow for increased spatial and
temporal resolutions for post-processing purposes, reduced I/O costs,
and significant decrease of time from data to insight.
This would therefore alleviate increasing difficulties encountered by
AMR simulation analysts caused by the current need to save a
sufficient amount of raw solution data to persistent storage.
\end{enumerate}
We acknowledge that some of these items are mutually independent,
and thus do not have to be executed in the order of the list.
However, [a] and [b] constitute the necessary foundation of the whole;
this article is therefore focused on these two first steps.

\section{Foundations}
\label{s:foundations}
Neither of the features for tree-based AMR grids, necessary per the
requirements detailed in~\S\ref{s:context}, were supported by
\VTK{} prior to 2012.
We therefore decided to design, and implement, what was then the first
of its kind support for such data sets. 
This preliminary work was released as a set of new classes in~\VTK; we
also briefly described its governing principles in~\cite{carrard:imr21}.
However, we never provided a comprehensive description of the
corresponding data objects, and how they relate to the class of AMR
meshes of interest in our applications.
In addition, several years have passed since this first approach to
the problem, and our ideas and implementations have matured and
solidified.
We therefore think that the time has come to provide an in-depth
exposition of the foundations necessary to achieve our vision outlined
above.

\subsection{Vertices, Graphs, and Trees}
It is beyond the scope of this article to provide an extensive picture
of graph nomenclature and classification; the interested reader can
refer, e.g., to~\cite{bondy:11} for a systematic treatment of the
theory of graphs.
The fundamental building blocks of our trees are \emph{vertices},
which can also be implemented as data objects containing various
quantities of interest, such as simulation data, and mesh topology or
geometry attributes.
Given a set of vertices $V$, we then define an \emph{undirected edge}
as a pair set of vertices, with the following requirements for the set
of all undirected edges:
\begin{enumerate}[(i)]
\item
be connected, i.e., any two vertices are connected by a path of
adjacent edges, and
\item
not contain any cycle, i.e., a set of edges forming a closed polygon.
\end{enumerate}
In addition, one (and only one) vertex is chosen in $V$ to be the
\emph{root}.
In this setting, the directed edges are immediately deduced from the
undirected ones with the implicit ordering based on distance from the
root, in the sense of number of edges needed to transitively connect
to it.
Note that this is implicit ordering is indeed unambiguous:
on one hand, thanks to the connectivity axiom (i), any vertex in
${V}$ can always be connected to the root with a finite subset of
undirected edges, called a \emph{path}.
Furthermore, this path is unique: otherwise, a plurality of such paths
would contradict the acylicity axiom (ii); one can then define a unique
\emph{depth}  as being the number of edges in this path.
Finally, at least one vertex does not have any directed edge leaving
it, and any vertex that has this property is called a \emph{leaf}; all
non-leaf vertices are called \emph{strict nodes}.

\begin{figure}
\centering
\includegraphics[width=0.75\columnwidth]{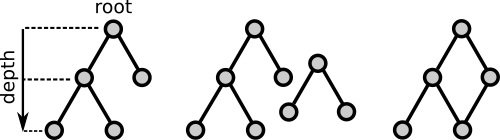}
\caption{Three different types of graphs: tree (left); not connected
graph (center), not a tree but nonetheless a \emph{forest}; and a
graph containing a cycle (right), therefore not a tree.
We decide to always show the root at the top.}
\label{fig:Trees-NonTrees}
\end{figure}
What matters most for a correct understanding of this article, is to
pay attention to the fact that typical usage of the term \emph{tree}
in Computer Science refers to a \emph{directed, rooted, acyclical
graph}, whereas in Mathematics it is more broadly understood as an
\emph{undirected, transitive, acyclical graph}.
The double $(V,E)$, where $E$ is the set of all directed
edges, is the definition of a \emph{tree} to be used thereafter.
A handful of examples and counter-examples are provided in
Figure~\ref{fig:Trees-NonTrees}; note that, for concision, we never
represent the directionality of the edges, for it is implicit
as we use the convention to always represent a tree
with its root at the top.
We also decide to always horizontally align vertices that have the
same depth.

\subsection{The Hyper Tree Object}
\label{s:hypertree}
We now introduce the concepts specific to our work, and in particular
the following, for which there are different definitions in the literature:
\begin{definition}
\label{def:hypertree}
A \emph{hyper tree object} (shorthand \emph{hypertree}) in
dimension~$d\in\mathbb{N}^*$ with \emph{branching factor}
$f\in\mathbb{N}^*$, is a type of data set that can be represented as a
tree, and where each strict node has exactly $f^d$ children.
In addition, primary attributes of this data set are attached to the
vertices of the tree.
\end{definition}
\begin{remark}
The range of AMR grids we want to support for our applications is
limited to the possible combinations of $d\in\{1;2;3\}$ and
$f\in\{2;3\}$.
The corresponding objects are called,
in dimension~$1$,
\emph{bintrees} ($f=2$) and \emph{tritrees} ($f=3$),
in dimension~$2$,
\emph{quadtrees} ($f=2$) and \emph{$9$-trees} ($f=3$), and
in dimension~$3$,
\emph{octrees} ($f=2$) and \emph{$27$-trees} ($f=3$).
\end{remark}

\begin{figure}[htb]
\centering
\begin{minipage}[t]{0.36\columnwidth}
\centering
\vspace{0pt}
\includegraphics[width=0.95\textwidth]{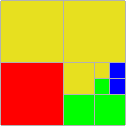}
\end{minipage}
\hfil
\begin{minipage}[t]{0.6\columnwidth}
\centering
\vspace{0pt}
\includegraphics[width=0.92\textwidth]{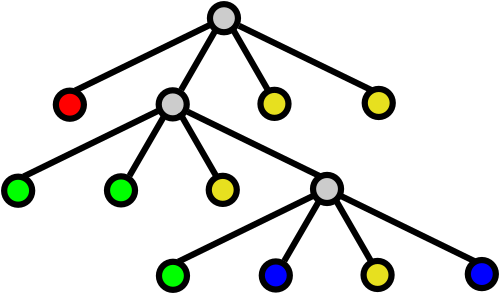}
\end{minipage}
\caption{Left: a $2$-dimensional AMR mesh obtained with 3 levels of
successive binary refinements of a quadrilateral; right: the
corresponding hypertree representation.
Colors are used to represent the attribute values attached to mesh
cells.}
\label{fig:HyperTreeBinary2D}
\end{figure}
There is a trivial bijection between hypertree objects and tree-based
AMR meshes descending from a unique root cell: for instance, each leaf
of a hypertree object $\mathcal{H}$  represents exactly one mesh cell
that is not refined, whereas strict nodes in $\mathcal{H}$ are
bijectively associated with all \emph{coarse cells} (i.e., cells in
the mesh that are subdivided).
This bijective construction is illustrated with the case of a quadtree
in Figure~\ref{fig:HyperTreeBinary2D}.
Note that this AMR mesh does not have attribute values attached to
coarse cells, whence the gray color in the corresponding strict tree
nodes.
However, it is possible to assign attribute values to strict nodes,
and in fact some CEA simulations codes compute attribute values at
coarse cells.
Furthermore, coarse cell attributes can be computed during
post-processing, e.g. for level-of-detail (LOD) purposes.
Last, as will be seen later in this article, we exploit this
capability to store attribute values at strict nodes in the aim of
optimizing tree traversals for some classes of filters.

Regarding the geometry of a hypertree object, this work addresses
the case of AMR meshes embedded in the $3$-dimensional Euclidean
space $\mathbb{R}^3$, irrespective of their actual dimensionality.
We thus do not provide support for higher dimension meshes, which are not
needed by current AMR simulation codes.
Another consequence is that we do not handle planar nor linear AMR
grids natively; rather, they are always viewed as a 3D object with
one or two fixed coordinates.
This might appear as a sub-optimal setting, which it is in some very
limited respect because, as we will see, coordinates do not have to be
stored for all cells.
But, on the other hand, \VTK{} is optimized for dimension 3, and its
rendering system does not provide a good way to co-mingle objects of
different dimensionalities in a generic fashion.
Furthermore, some of the visualization filters considered, and in some
cases developed, produce AMR outputs that have lower dimensionality
than the AMR inputs.
Therefore, this trade-off appears to be the best one under the current
circumstances: choice of the visualization library as well as
range of potential data sets.

Because the considered AMR meshes are always rectilinear, the geometry
of a hypertree object is implicitly but nonetheless unambiguously
specified, given:
\begin{enumerate}
\item the origin $\overrightarrow{x}=(x_0;x_1;x_2)\in\mathbb{R}^3$ of
the root node;
\item the size $\overrightarrow{s}=(s_0;s_1;s_2)\in\mathbb{R}^3$ of
the root node; and
\item the direction (resp. normal, first axis) vector
$\overrightarrow{v}\in\mathbb{R}^3$ of the root node in
dimension~$1$ (resp. $2$, $3$).
\end{enumerate}
We made the additional design choice to natively support only
axis-aligned tree-based AMR meshes, because \VTK{} has a generic
approach to supporting geometric transformations such has
translations, rotations, and homothecies.
This allows us to use an \emph{orientation} value 
$o\in\mathbb{N}$ only, in order to specify the direction of the axis
along which a 1-dimensional AMR mesh is aligned, or the normal to the
plane inside which a 2-dimension mesh is contained, in order to
specify the embedding into $\mathbb{R}^3$.
If $d=3$, the convention is that $o=0$ as this value is not used anyway.
Last, thanks to the size vector $\overrightarrow{s}$, supported AMR
geometries are therefore not limited to unit segments, squares and
cubes, but also include arbitrary segment lengths and rectangular
shapes.
Using these notations, the triple
$(\overrightarrow{x};\overrightarrow{s};o)\in(\mathbb{R}^3)^2\times\mathbb{N}$
is called the \emph{$3$-dimensional embedding of} the
considered hyper tree.

\subsection{Mapping and Indexing}
\label{s:mapping-indexing}
No particular indexing is assumed in tree structures in general, for
it depends on the particular \emph{traversal scheme} utilized to visit
tree vertices and index them accordingly.
However, we need to have a consistent mapping scheme, in order to
unambiguously map AMR meshes in the class we want to
address, into hypertree objects.
In Figure~\ref{fig:HyperTreeBinary2D} for instance, implicit
orderings of mesh cells and of tree vertices were used, in order to
map one into the other.
It is natural, following the flow in which mesh refinement is
performed, to convention that mesh cells as well as tree vertices
are ordered by depth, root first, which is another way of saying that both mesh
and tree are traversed by \emph{breadth-first search}
(BFS)~\cite{bondy:11} traversal.

Meanwhile, because there is no unique order over
$\llbracket0;f\llbracket^d$, as soon as $d>1$, it is easy to convince
oneself that choosing a particular BFS scheme amounts to determining a
unique way to locally index the sub-cells obtained by refining a coarse
cell $C$, called \emph{children cells of}~$C$.
By definition, there are always $f^d$ such children cells to any coarse
cells, which can each be uniquely identified in terms of the number of
refinements at which they begin along each axis.
The corresponding indices in $\llbracket0;f\llbracket^d$, shown in
Figure~\ref{fig:BranchingFactors2D} in the case where $d=f=2$, are
called the \emph{child coordinates}.
In order to generalize the construction illustrated in this example, to
map an arbitrary AMR mesh into a hypertree, in an unambiguous manner,
one needs to define a particular bijection from the $d$-dimensional
Cartesian product $\llbracket0;f\llbracket^d$ into the ordered set
$\llbracket0;f^d\llbracket$.
For all our work, we decide to use the following convention:
\begin{definition}
The \emph{hypertree child index map} $\Upphi_{d,f}$,
with~$(d,f)\in{\mathbb{N}^*}^2$ is the lexicographic
order over $\llbracket0;f\llbracket^d$.
\label{def:child-index}
\end{definition}
It is beyond the scope of this article to discuss the lexicographic
order over Cartesian products in a detailed way; suffices to know that
it is the analog to the lexicographic order over finite words in a
finite alphabet (the dictionary order) and that it indeed provides a
total order.
In addition, one has the following property, whose proof is left to
the interested reader as an exercise (by recurrence over $d$):
\begin{proposition}
\label{pro:child-index-map}
Given child coordinates $(c_0,\dots,c_{d-1})$ in dimension~$d$:
\[
\Upphi_{d,f}(c_0,\dots,c_{d-1}) = \sum_{k=0}^{d-1} c_k f^k.
\]
\end{proposition}

\begin{figure}
\centering
\centering
\begin{minipage}[t]{0.42\columnwidth}
\centering
\vspace{0pt}
\[
\renewcommand{\arraystretch}{1.1}
\setlength{\arraycolsep}{1mm}
\begin{array}{@{}r|cc@{}}
      1 & 6 & 7 \\
      0 & 4 & 5 \\\hline
c_2 = 1 & 0 & 1 
\end{array}
\]
\[
\begin{array}{@{}r|cc@{}}
      1 & 2 & 3 \\
      0 & 0 & 1 \\\hline
c_2 = 0 & 0 & 1 
\end{array}
\]
\end{minipage}
\begin{minipage}[t]{0.55\columnwidth}
\centering
\vspace{8pt}
\includegraphics[height=28mm]{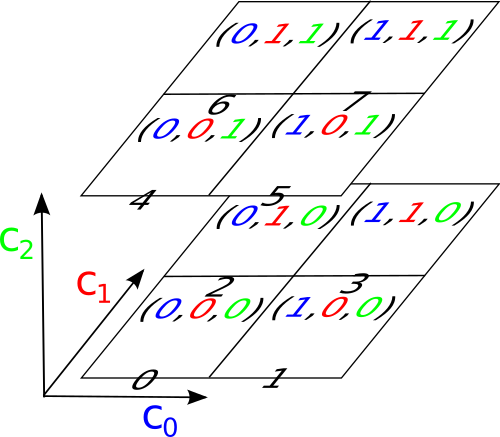}
\end{minipage}
\hfil
\caption{The hypertree child index map in the $3$-dimensional binary
case ($d=3$, $f=2$).}
\label{fig:Mapping3D-Binary}
\end{figure}
The index maps for $d=1$ are simply the identities over
$\llbracket0;f\llbracket$.
For the values of $d$ and $f$ that are of practical interest to us,
the $(c_0,c_1,c_2)$ tables are given in
Figure~\ref{fig:Mapping3D-Binary} for $f=2$, and by the following
tables for $f=3$:
\[
\renewcommand{\arraystretch}{1.1}
\setlength{\arraycolsep}{1mm}
\begin{array}{@{}r|ccc@{}}
      2 & 6  & 7  & 8 \\
      1 & 3  & 4  & 5 \\
      0 & 0  & 1  & 2 \\\hline
c_2 = 0 & 0  & 1  & 2 
\end{array}
\quad
\begin{array}{@{}r|ccc@{}}
      2 & 15 & 16 & 17 \\
      1 & 12 & 13 & 14 \\
      0 & 9  & 10 & 11 \\\hline
c_2 = 1 & 0  & 1  & 2 
\end{array}
\quad
\begin{array}{@{}r|ccc@{}}
      2 & 24 & 25 & 26 \\
      1 & 21 & 22 & 23 \\
      0 & 18 & 19 & 20 \\\hline
c_2 = 2 & 0 & 1 & 2 
\end{array}
\]
When considering only the 2 tables with $c_2=0$ amongst the above, one
obtains the corresponding maps for $d=2$.
\subsection{The Hyper Tree Grid Object}
In order to account for a broad category of tree-based AMR grids,
including those that do not have uniform geometry along each axis,
or whose initial refinement pattern is not that of hypertree, we
introduced a broader-scoped object in 2012, which have since deeply
modified and are discussing fully now.
\begin{figure}[h]
\centering
\includegraphics[width=0.8\columnwidth]{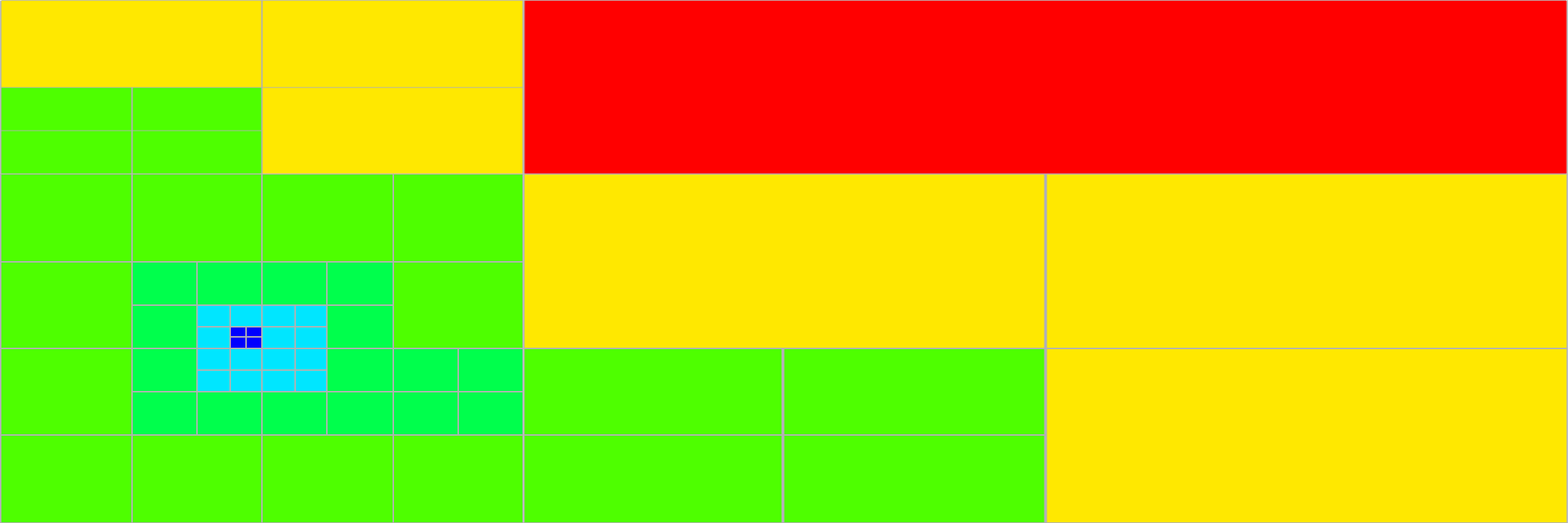}
\caption{A $2$-dimensional AMR mesh obtained with 4 levels of
successive binary refinements of $3\times2$ rectilinearly aligned
hypertree objects with different sizes along each axis.}
\label{fig:HyperTreeGridBinary2D}
\end{figure}
\begin{definition}
Let $\mathcal{H}$ and $\mathcal{H}'$ be two hypertree objects, with
same dimension~$d\in\{1;2;3\}$ and \emph{branching factor}
$f\in\mathbb{N}^*$, with respective $3$-dimensional embeddings
$(\overrightarrow{x};\overrightarrow{s};o)$ and
$(\overrightarrow{x}';\overrightarrow{s}';o')$.
If
\[
\exists k\in\{0;1;2\} \quad
\renewcommand{\arraystretch}{1.2}
\left\{\begin{array}{l}
x_k' = x_k + s_k \\
\forall l\in\{0;1;2\}\setminus{\{k\}} \; (x_l',s_l') = (x_l,s_l)
\end{array}\right.
\]
and, when $d\neq3$, $o'=o$, we say that $\mathcal{H}'$ is
\emph{rectilinearly consecutive to} $\mathcal{H}$
\emph{for component $k$}, denoted
$\mathcal{H}\underset{k}{\prec}\mathcal{H}'$.
\end{definition}
Intuitively, what this means is that the outside boundary of
$\mathcal{H}\cup\mathcal{H}'$ is a line segment in dimension~$1$, a
rectangle in dimension~$2$, and a rectangular prism 
in dimension~$3$, with origin and orientation equal to those of
$\mathcal{H}$, and size vector as well, except for its $k$ component
which is equal to $\overrightarrow{s_k}+\overrightarrow{s_k}'$.
For example, Figure~\ref{fig:HyperTreeGridBinary2D}, are shown $6$
binary hypertree objects in dimension~$2$, arranged in to have
rectilinear consecutiveness for components~$0$ and~$1$. 

Given any triple $t\in\mathbb{N}^3$, we denote $\Uppi{t}$ the
product of its components, and $\llbracket{t}\llbracket$ the set of
triples $t'\in\mathbb{N}^3$ such that $t'<t$ in the lexicographic 
sense. For example,
\[
\llbracket{(3;2;2)}\llbracket = 
\big\{\{0;0;0\};\{0;0;1\};\{0;1;0\};\dots;\{2;1;1\}\big\}.
\]
We now introduce our main object:
\begin{definition}
\label{def:hypertreegrid}
A \emph{hyper tree grid object} (shorthand \emph{hypertree grid}) in
dimension~$d\in\{1;2;3\}$ with \emph{branching factor}
$f\in\mathbb{N}^*$ and \emph{extent}
$E\in{\mathbb{N}^*}^d\times\{1\}^{3-d}$, denoted
$\mathcal{G}_E^{d,f}$, is a type
of data set comprising $\Uppi{E}$ hyper tree objects
in dimension~$d$ and with branching factor~$f$,
denoted $\mathcal{H}_{i,j,k}$ where $(i;j;k)\in\llbracket{E}\llbracket$,
such that
\[
\forall(i;j;k)\in\llbracket{E}\llbracket\;
\left\{\begin{array}{lcl}
i+1<E_0 & \Rightarrow
& \mathcal{H}_{i,j,k} \underset{0}\prec \mathcal{H}_{i+1,j,k}\\
j+1<E_1 & \Rightarrow
& \mathcal{H}_{i,j,k} \underset{1}\prec \mathcal{H}_{i,j+1,k}\\
k+1<E_2 & \Rightarrow
& \mathcal{H}_{i,j,k} \underset{2}\prec \mathcal{H}_{i,j,k+1}
\end{array}\right.
\]
In addition, primary attributes of this data set are attached to the
individual hypertrees.
\end{definition}
Given an arbitrary hyper tree grid object $\mathcal{G}_E^{d,f}$, we 
denote $\mathcal{H}_{i,j,k}^{d,f}$ the hyper tree object with
discrete coordinates $(i;j;k)\in\llbracket{E}\llbracket$ and call it
the \emph{constituting hypertree of} $\mathcal{H}_E^{d,f}$ \emph{at
position} $(i;j;k)$.
Under the assumptions of Definition~\ref{def:hypertreegrid} regarding
$d$, $f$, and $E$, we have:
\begin{proposition}
\label{pro:hypertreegrid-embedding}
The outside boundary of a hyper tree grid object $\mathcal{G}_E^{d,f}$ is a
$d$-dimensional rectangular prism, uniquely determined by the
$3$-dimensional embeddings of its constituting hypertree objects.
\end{proposition}
\proof
Without loss of generality, it is sufficient to prove this assertion
in dimension~$3$: the result in dimension~$2$ (resp.~$1$) ensues by
setting $E_2=1$ (resp. $E_1=E_2=1$).
By definition, given any 2 integers $a$ and $b$ such that
$(1,a,b)\in\llbracket{E}\llbracket$,
$\mathcal{H}_{0,a,b}^{d,f}\underset{0}{\prec}\mathcal{H}_{1,a,b}^{d,f}$
and, for all $i\in\llbracket0;E_0-1\llbracket$, 
$\mathcal{H}_{i,a,b}^{d,f}\underset{0}{\prec}\mathcal{H}_{i+1,a,b}^{d,f}$.
Therefore, by recurrence, the outside boundary $R_{a,b}$ of
$\cup_{i<E_0}\mathcal{H}_{i,a,b}$ is a rectangular prism 
with origin and size equal to those of $\mathcal{H}_{0,a,b}$, with the
exception of the first component of its size vector, that is equal to
the sum of the first components of the size vectors along the first
axis.

Applying the same argument to such stacks, with the form
$\cup_{i<E_0}\mathcal{H}_{i,a,b}$, that are consecutive along the
second axis, one obtains that the outside boundary $R_b = \cup_a R_{a,b}$ of
$\cup_{i<E_0,j<E_1}\mathcal{H}_{i,j,b}$ is also a rectangular
prism, with origin and size equal to those of $\mathcal{H}_{0,0,b}$,
with the exception of the first and second components of its size
vector, that are equal to the sums of the first and second components
of the size vectors along the first and second axes, respectively.

Finally, stacking consecutive blocks
$\cup_{i<E_0,j<E_1}\mathcal{H}_{i,j,b}$ that are consecutive
along the third axis, one finally obtains the outside boundary
$R=\cup_b R_b$ of $\mathcal{G}_E^{d,f}$, a rectangular
prism, with origin equal to that of $\mathcal{H}_{0,0,0}$,
and with size vector whose components are the sums of the
corresponding components of the size vectors along the 3 axes,
respectively.
\qed\\
\begin{remark}
\label{rem:rectilinear-coordinates}
Applying the same argument to the origin vectors of the constituting
hypertrees shows that these are exactly the vertex coordinates of a
rectilinear grid, whose elements are exactly the bounding boxes of
said hypertrees.
It is therefore not necessary to \emph{explicitly} store the $\Uppi{E}$
geometric embedding triples (i.e., $2d\Uppi{E}$ floats and $\Uppi{E}$
integers) to describe the geometry of the hypertree grid.
Rather, it is sufficient to describe it \emph{implicitly} by storing
one coordinate array per dimension and a single orientation for the
entire grid, at a much smaller total cost between
$d(\sqrt[d]{\Uppi{E}}+1)$ (best case: $\mathcal{G}_{(a,a,a)}^{3,f}$)
and $\Uppi{E}+5$ (worst case: $\mathcal{G}_{(a,1,1)}^{d,f}$) all
hypertrees consecutive along a single direction) floats, plus a single
integer. 
\end{remark}

Finally, a map providing a direct look-up from the \emph{local index}
$n_{\mathrm{l}}$ of a vertex inside the constituting hypertree
$(i;j;k)\in\mathbb{N}^3$, into the \emph{global index}
$n_{\mathrm{g}}$ of this vertex relative to $\mathcal{G}_E^{d,f}$ as a
whole, is needed in order to retrieve attribute values at any given
vertex.
Such a map thus must take the following form:
\begin{definition}
\label{def:global-index}
The \emph{global index map} $\Upgamma_{d,f,E}$ of a hypertree grid
$\mathcal{G}_E^{d,f}$ is an injective map:
\[
\begin{array}{rccl}
\Upgamma_{d,f,E}: 
& \mathbb{N}^4 & \longrightarrow & \mathbb{N} \\
& (n_{\mathrm{l}};i;j;k) & \longmapsto & n_{\mathrm{g}}.
\end{array}
\]
\end{definition}

We hereafter freely identify any rectilinear,
tree-based AMR meshes with its corresponding hypertree grid object,
referring to this process as that of \emph{identification}, so
that it not be confused with that of \emph{duality} which we are now
going to discuss.
\subsection{The Dual Mesh}
\label{s:dual}
In order to resolve the difficulties posed by AMR T-junctions, two
different approaches are possible: one is to implement new processing
algorithms specialized towards tree-based AMR grids, the other
consists of transforming the AMR input into a conforming unstructured
grid, allowing for reuse of existing algorithms.
In earlier work~\cite{carrard:imr21} (which, in our knowledge, is the
only existing work in this field), we chose the latter approach, by
the means of defining a \emph{dual grid} construction, upon which all
filters designed for vertex-centered attributes (i.e., iso-contouring)
could natively operate when all variables are cell-centered.
In this case, the visualization results are correct, provided the
attributes correspond to variables values computed at
cell centers -- as opposed to averaged over the cell.
That was a strong limitation of this approach, from its inception, as
most cell-centered simulations use the latter rather than the
former.

Nonetheless, this notion of duality remains a powerful conceptual tool
when the considered visualization technique requires that the
elements of a conforming mesh be generated, used, and disposed of,
one at a time (as opposed to creating the entire mesh)  
In what follows, by \emph{mesh} we refer to a polytopal cover of a
finite, closed subset of an Euclidean space.
The reader interested in an in-depth discussion of meshes, and how
they relate to numerical simulations, can refer in particular
to~\cite{frey:08}.
\begin{definition}
Given $d\in\{1;2;3\}$ and a $d$-dimensional mesh $\mathcal{M}$,
referred to as the \emph{primary mesh}, we define its \emph{dual mesh}
$\mathcal{M}^\ast$ as follows:
\begin{enumerate}[(i)]
\item
to every $d$-dimensional cell $e\in\mathcal{M}$ is associated a
\emph{dual vertex} $e^\ast\in\mathcal{M}^\ast$, with coordinates those
the isobarycenter of~$e$;
\item
to every vertex $v\in\mathcal{M}$ is associated a \emph{dual cell}
$v^\ast\in\mathcal{M}^\ast$, whose vertices are exactly the dual
vertices $e_i^\ast\in\mathcal{M}^\ast$ such that $v$ is a vertex of $e_i$.
\end{enumerate}
Depending on the value of $d$, \emph{dual edges} and \emph{dual faces}
are defined as the $1$ and $2$ dimensional elements of the dual cells,
respectively.
\end{definition}
\begin{remark}
Note that this definition is not that of duality in the
Delaunay-Vorono\"{\i} sense~\cite{delaunay:34} because, as a result
of the finite extent of $\mathcal{M}$, there is no bijection between
the vertices of $\mathcal{M}$ and the cells of $\mathcal{M}^\ast$
although, by construction, there is a bijection between the cells of
$\mathcal{M}$ and the vertices of $\mathcal{M}^\ast$.
\end{remark}

\begin{figure}[h]
\centering
\begin{minipage}[t]{0.49\columnwidth}
\centering
\vspace{0pt}
\includegraphics[height=35mm]{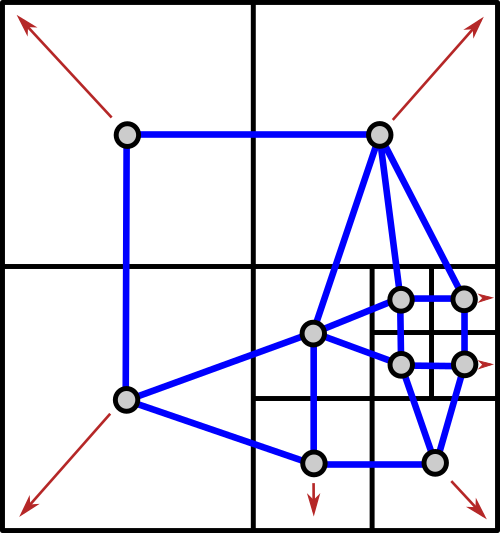}
\end{minipage}
\hfil
\begin{minipage}[t]{0.49\columnwidth}
\centering
\vspace{0pt}
\includegraphics[height=35mm]{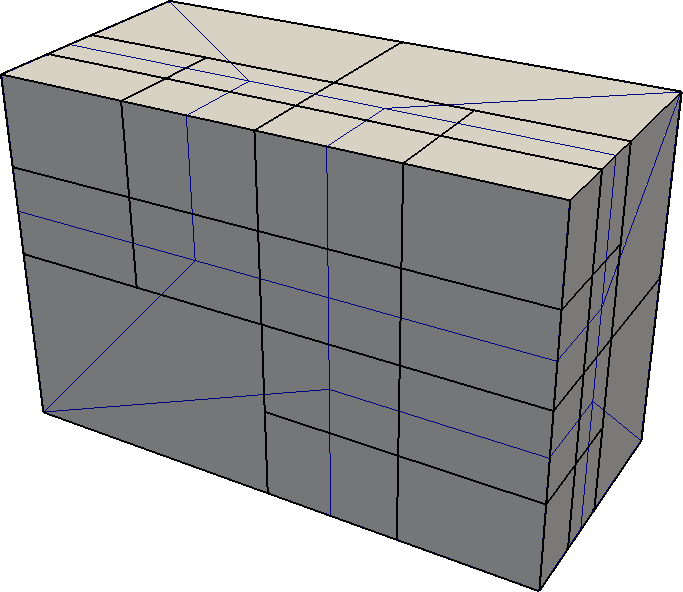}
\end{minipage}
\caption{Left: a 2D ternary tree-based AMR mesh (in black), overlaid
with its dual (in blue); arrows indicate how the dual vertices may
be displaced to produce an adjusted dual.
Right: a 3D binary tree-based AMR grid (in black)
overlaid with its adjusted dual (in blue).}
\label{fig:PrimalAndDual2D3D}
\end{figure}

When applied to the case of AMR meshes, this definition must be
understood in the sense that~$\mathcal{M}$ only contains the
non-refined cells (i.e. hypertree leaves), excluding all coarse cells
(i.e. hypertree strict nodes).
Furthermore, we often also use an \emph{adjusted dual}
$\varphi(\mathcal{M}^\ast)$, where $\varphi$ is a geometric
transformation that maps the vertices belonging to
$\partial\mathcal{M}^\ast$, the boundary of $\mathcal{M}^\ast$, so
that
\[
\partial\varphi(\mathcal{M}^\ast) = \partial\mathcal{M}.
\]
Note that there is no unicity in the choice of $\varphi$; 
examples of this construction are provided, in dimension~$2$ and ~$3$,
in Figure~\ref{fig:PrimalAndDual2D3D}.
Using the notion of \emph{conforming mesh} in the
sense of a polytopal covering where two $k$-dimensional items are
either distinct or their intersection is exactly a shared
$(k-1)$-dimensional item of the mesh, we have the following key
result, whose proof is left to the reader as an exercise:
\begin{proposition}
\label{pro:dual-conforming}
If $\mathcal{M}$ is formed by the refined cells of a tree-based
rectilinear AMR mesh, then $\mathcal{M}^\ast$ is a conforming mesh.
\end{proposition}
By definition, a T-junction is the topological configuration where two
edges (i.e. $1$-dimensional mesh items) intersect at a vertex (i.e. a
$0$-dimensional mesh item) which is not shared by both edges.
Therefore, if a mesh has one T-junction, it is not conforming; by
contraposition of Proposition~\ref{pro:dual-conforming}, it follows
that the problem of T-junctions as illustrated, e.g. in
Figure~\ref{fig:AMR-Unstructured-2D}, vanishes when replacing
the primal AMR mesh with its dual or its adjusted dual (which does not
modify the topology).

\subsection{Cursors and Supercursors}
\label{s:cursors-supercursors}
We are now at a point where we can represent and store all
tree-based, rectilinear AMR meshes we want to support.
The question that immediately follows is that of operating on these,
by means of appropriately designed \emph{hypertree grid filters}.
In order to do this, such filters must be endowed with an efficient
way to both access and traverse hypertree grid objects.
Because a hypertree grid $\mathcal{G}$ is inherently a list of
hypertrees, it is only natural to iterate over these as a way to
traverse $\mathcal{G}$ in its entirety.
In general, \emph{depth-first search} (DFS) traversal of trees is
efficient, and is therefore our preferred \emph{modus operandi},
whenever no other order of traversal is explicitly needed.
Meanwhile, an algorithm that needs to iterate over all vertices of
$\mathcal{G}$ will also typically need access to some of
the information contained there, such as global indices
allowing for attribute retrieval from data arrays.
We thus introduce the following object:
\begin{definition}
A \emph{hypertree cursor} is a structure pointing to a hypertree,
that can both traverse it and access its vertex attributes.
\end{definition}
A minimal hypertree cursor will therefore comprise a
reference to the underlying hypertree, together with a
stack-like data structure storing the path from the root to the
\emph{current vertex}, i.e. the vertex towards which the cursor is
pointing.
It will also be endowed with at least the two following operators:
\begin{description}
\item[\texttt{ToParent()}:]
move one vertex up in the tree, except if already at the root.
\item[\texttt{ToChild(i)}:]
descend into the vertex with child index \texttt{i}
(cf. Definition~\ref{def:child-index}), relative to the vertex
currently pointed at, except if already at a leaf.
\end{description}
Any actual implementation will also equip this structure with other
operators such as data accessors, direct or indirect, to
vertex attributes.
Because the AMR meshes we want to address only have cell-wise
attributes, the corresponding data arrays are all of equal length and
allow for random access per cell index.
In order to keep the hypertree cursor as lightweight as possible, we
chose to provide only indirect access to attribute values, which can
be achieved with a single instance variable storing vertex, and
therefore mesh cell, indices into the corresponding data arrays.
This turns into the requirement that attribute fields be all ordered
in the same fashion, using the global indexing scheme
(cf. Definition~\ref{def:global-index}).

\begin{figure}[h]
\centering
\includegraphics[width=0.65\columnwidth]{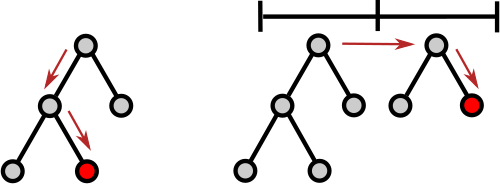}
\caption{DFS traversal to the red-colored leaf, in a hypertree (left)
and a hypertree grid (right).}
\label{fig:HyperTreeCursor-HyperTreeGridCursor}
\end{figure}
However, because individual hypertrees are never used on their
own but, rather, are always interlocked in a broader hypertree
grid~$\mathcal{G}$, the hypertree cursor is not sufficient to
traverse~$\mathcal{G}$. 
Rather, it must be enriched with additional topological information,
allowing it to move from one tree root to the next, as if there were
a meta-root vertex, from which all hypertree roots would descend.
This means, in particular, that both \texttt{ToParent()} and 
\texttt{ToChild()} must be equipped with additional logic, in order to
properly traverse across level $0$ cells.
This is illustrated in
Figure~\ref{fig:HyperTreeCursor-HyperTreeGridCursor}, showing the 
path taken by a hypertree cursor searching a vertex with a
give attribute value, compared to that of a hypertree grid cursor
doing the same inside a $2\times1\times1$ hypertree grid.

A hypertree cursor, extended to allow for traversal across a grid
of hypertree objects, is naturally called a \emph{hypertree grid
cursor}.

Furthermore, many visualization filters require neighborhood
information to perform their computations.
For example, an outside boundary extraction filer, in dimension~$3$,
needs to know whether a given cell has neighbors across any of its
faces, as a boundary face is generated if and only if it is not
shared by two cells.
In order to provide neighborhood information we devised and implemented the
following compound structures:
\begin{definition}
A \emph{supercursor} is a hypertree grid cursor keeping track of a
neighborhood of cursors while traversing the hypertree grid.
\end{definition}
\begin{figure}[h]
\centering
\includegraphics[width=0.9\columnwidth]{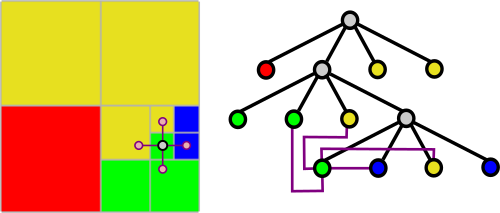}
\caption{Left: $4$-neighborhood of a cell in a $3$-deep, $2$-dimensional binary AMR
mesh; right: same neighborhood when mapped
to the hypertree representation of the mesh.} 
\label{fig:HyperTreeBinary2D-Supercursor-VonNeumann}
\end{figure}
For the sake of simplicity, let us consider the case of a hypertree
grid containing a single hypertree.
One such example is readily provided by the single-root AMR mesh of
Figure~\ref{fig:HyperTreeBinary2D}, left.
Let us then turn our attention to the single green cell that is to be
found at the deepest refinement level: it has $4$ neighbor across its
edges, marked by the cross-shaped structure in
Figure~\ref{fig:HyperTreeBinary2D-Supercursor-VonNeumann}, left.
This same structure is then showed, to the right, after having been
mapped onto the hypertree equivalent of the AMR mesh: the set of
purple multi-lines connecting the green leaf at maximum depth to 4 other tree
leaves is, effectively, the supercursor state when it points to the
green leaf with depth~3.
Implementing this supercursor thus entails providing the logic
necessary to update these links, when moving vertically within a hypertree,
as well as when moving horizontally from one hypertree root to a
neighboring one in the Cartesian grid of roots.
\begin{remark}
It is important to note that a supercursor is not a hybrid DFS/BFS
traversal structure:
it can only retrieve information from the vertices to which it is
linked, but it cannot directly traverse to them.
The complexity and computational cost of a traversal object able to
achieve that end would be prohibitive indeed.
\end{remark}

\begin{figure}[ht]
\centering
\includegraphics[width=0.9\columnwidth]{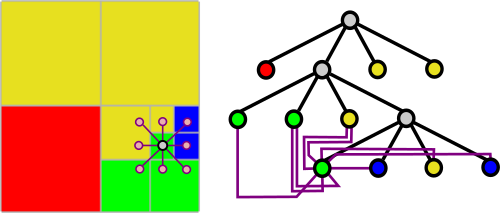}
\caption{Left: $8$-neighborhood of a cell in a $3$-deep, $2$-dimensional binary AMR
mesh; right: same neighborhood when mapped
to the hypertree representation of the mesh.} 
\label{fig:HyperTreeBinary2D-Supercursor-Moore}
\end{figure}
Another complication arises from the fact that the notion of
\emph{neighborhood} itself is not invariant, but rather depends on the
considered topology.
In our case, choosing a topology amounts to defining a criterion to
decide whether $2$ cells in an AMR mesh are \emph{connected}.
For instance, the already introduced hypertree and hypertree grid
cursors are supercursors with respect to the discrete topology.
In dimension~$1$ (resp. $2$, $3$), another type of connectivity is
called the $2$- (resp. $4$-, $6$-) connectivity, where two cells
are connected if and only if their share a common vertex (resp., edge,
face).
This type of connectivity defines the corresponding
\emph{$d$-dimensional Von Neumann neighborhood}~\cite{packard:85}.

Moreover, some algorithms need richer neighborhoods, expanding this
criterion to include connectivity across mesh vertices for
$d=2$, and also across mesh edges for $d=3$.
For instance, in order to compute a dual cell associated
with a given vertex $v$ of a primal cell $C$ in dimension~$2$, it is
necessary to iterate over all mesh cells that share $v$ with $C$,
i.e., all those that are connected to $C$ either by an edge that 
contains $v$, or by $v$ alone.
A Von Neumann neighborhood no longer suffices for this purposes.
In dimension~$3$, the problem is further compounded by the distinction
between per-vertex, per-edge, and per-face connectivity.
This defines the corresponding \emph{$d$-dimensional Moore
neighborhood}.
Figure~\ref{fig:HyperTreeBinary2D-Supercursor-Moore} illustrates the
difference between these two neighborhood types in dimension~$2$, with
the Moore supercursor pointing to $3^d-1=8$ neighbors, of which only
$2d=4$ belong to the Von Neumann neighborhood of
Figure~\ref{fig:HyperTreeBinary2D-Supercursor-VonNeumann}.

Note that, in dimension~$1$, there is not distinction between von
Neumann and Moore connectivities.
Furthermore, in dimension~$2$, these two are distinct but without any
intermediate in the scale of connectivity pattern whereas, in
dimension~$3$ one could also consider \emph{18-connectivity}, i.e.,
where two mesh cells are connected if and only if they share a face or
an edge (but a vertex only is not sufficient).
However, we have not found so far any use case where this type of
connectivity would be useful.
Other types of neighborhoods could also be defined, e.g., with
vicinity stencils spanning more that one cell on each side, as
required by the considered algorithm.

\begin{figure}[ht]
\centering
\begin{minipage}[t]{0.9\columnwidth}
\includegraphics[width=\columnwidth]{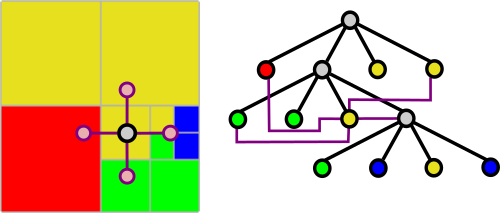}\\
\includegraphics[width=\columnwidth]{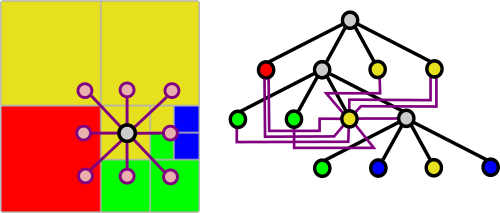}
\end{minipage}
\caption{Von Neumann (top) and Moore (bottom) neighborhoods
of a cell that has neighbors with greater depth,
in a $3$-deep, $2$-dimensional binary AMR mesh.}
\label{fig:HyperTreeBinary2D-SupercursorV2}
\end{figure}
However, the notion of vicinity of a cell is not always unambiguous.
Specifically, given a cell $C$ at depth $\delta(C)$ in the mesh and
one of its lower-dimensional entities $e$, define $\mathcal{N}(C,e)$
the set of all cells that are neighbors of $C$ across $e$ for the
considered neighborhood type.
When $\mathcal{N}(C,e)\neq\varnothing$, let $C'\in\mathcal{N}(C,e)$ that
has greatest depth in $\mathcal{N}(C,e)$. This cell $C'$ is not
necessarily unique, and only the following two cases thus may occur:
\begin{enumerate}[(i)]
\item
$\delta(C')>\delta(C)$; in this case, $C'$ is a smaller cell than $C$ 
and other cells with the same size may share entity $e$ with $C$ as
well.
The neighbor cell to $C$ is thus chosen to be the unique
$C''$ in $\mathcal{N}(C,e)$ that has the same depth as $C$; this is
illustrated in Figure~\ref{fig:HyperTreeBinary2D-SupercursorV2}.
\item
$\delta(C')\le\delta(C)$; in this case, $C'$ is a cell at least as
large as $C$ and thus there can be no ambiguity for no other mesh
cells may share $e$ with $C$ as well. In this case, and only in this
case, $C'$ is chosen to be the neighbor of $C$ in the supercursor.
\end{enumerate}
Two important consequences result from the above: first, one same cell
can appear more than once in the vicinity cursors; second, a neighbor
of a leaf can be a coarse cell.

\subsection{The Material Mask}
\label{s:material-mask}
Further complexity arises from the fact that AMR simulations results
we wish to support can also distinguish between different
\emph{materials} participating in the simulation.
We now focus on how we have enriched the hypertree grid object in
order to both support this additional feature and at the same time
take advantage of it even in the absence of a material specification
to increase execution speed of some filters.

One first consideration is that the material properties are specified
on a per-cell basis, for coarse as well as for refined cells.
Furthermore, some simulations allow for the presence of
two different materials in a same cell, hereby implying the existence
of a \emph{material interface}, which can be approximated using
various techniques not discussed here.
By design, a coarse cell in the AMR mesh exists if and only it has at
least one leaf in its descent that contains the selected material.
Although this can result in an incomplete AMR mesh, when compared to
the whole computational domain, this trade-off is warranted by the
fact that the analyst generally knows which materials are strictly
necessary, and which ones can be ignored (e.g.,
vacuum, inert materials, etc.).
 
Our approach to handling the material, is to define an additional
cell-wise attribute, call the \emph{material mask}.
In practice, this is a bit array, sized as the other attribute arrays
of the considered hypertree grid, i.e., with a number of
entries that are equal to the number of vertices (both strict nodes
and leaves).

Our approach to processing hypertree grids with non-void material
mask is to consider masked vertices, as well as all their descent, as
being non-existent.
As a result, a DFS traversal will stop its descent as soon as a
masked tree vertex is encountered, irrespective of the fact that its
parent cell is \emph{stricto sensu} a strict node, hereby giving rise
to a notion of \emph{lato sensu} leaf: for all processing purposes, a
strict vertex in the hypertree, whose children are all masked, is
considered as a leaf node.
Moreover, one masked tree vertex hides to processing algorithm the
entirety of the sub-tree that descends from it.
Nonetheless, the entirety of the hypertree grid remains represented.

This design constraint, arising from the very nature of the notion
of material in an AMR mesh, must not only be accommodated, but can
also be abused, by being used as a form of \emph{virtual decimation}.
This results in dramatically accelerating execution speed of hypertree
grid filters, as they traverse an input mesh and obey
some logic to decide whether an input cell with generate
output items, or not.
In other words, for the sake of performance, a material mask can
be used to produce virtually decimated hypertree grids, at
the expense of the additional memory required to store the hidden
parts.
In general, a material mask is an attribute with negligible relative
memory footprint, for it only consumes one bit per tree vertex.
However, the additional memory space required by the virtually
decimated -- but nonetheless truly represented -- vertices might
become prohibitive.
An ideal use of this concept would therefore balance those two
relative costs in an adaptive fashion and decide when \emph{actual
decimation} shall be executed.

\section{Method}
\label{s:method}
After having established the necessary foundations for our work, we
now discuss the methodology that we used to turn this
theoretical framework into an actual implementation.
While~\S\ref{s:foundations} can be understood as a frame of
reference that shall not evolve much in the future, the concrete
methods discussed below are, by nature, subject to further
improvements or revisions.
In particular, the techniques which are discussed hereafter are
already improvements upon earlier versions:
we have completely revised our approach to utilizing the
dual, as well as the design of supercursors, with respect to our
our earlier presentation~\cite{carrard:imr21}.
We begin by describing our methodological choices for efficient
representation and indexing of hypertrees and hypertree grids.

\subsection{The Compact Representation}
\label{s:compact-representation}
The ratio between the number of strict nodes to the total number
of nodes remain within
$[\tfrac{1}{1+f^d};\tfrac{1}{f^d}]$, an tends towards the upper bound
of this interval as the number of nodes increases.
It thus follows that the number of leaves dominates that of strict
nodes. Which is why we sought to implicitly define the leaves,
while explicitly storing in memory only the strict nodes.
Such a \emph{compact representation} still allows for traversal,
at the cost of minimal additional processing when visiting the leaves
due to their implicitness compensated by fewer cache missing.

In order to fully describe a hypertree, it is therefore sufficient
that each strict node store one index to refer the first amongst
its children, called the \emph{eldest} child node.
It is indeed sufficient to only store a reference to the eldest
when all children of a given cell are created as once as a block of
contiguous indices, at the end of the node array, instead of
allocating memory for each child which might be non-leaf cell itself.
Furthermore, the size of this block is constant and equal to $f^d$, by
definition of a hypertree.
In addition, because all of these children, child leaf or child strict node, have the same parent, it
suffices that only the eldest child store its parent index.
In fact, in order to retrieve the parent of any given node,
only the extra step of finding the position of its eldest
sibling is thus necessary.

\begin{figure}[h]
\centering
\includegraphics[width=0.8\columnwidth]{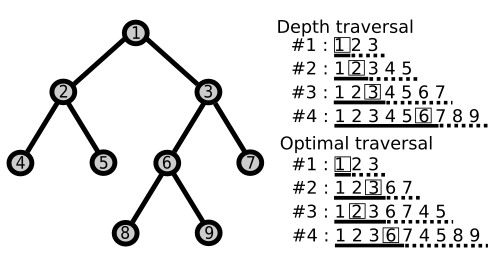}
\caption{Construction of a binary hypertree in dimension~$1$: the
order in which it is performed impacts the total memory footprint.
At each step ({\scriptsize \#}), the index in a square is that of the
nodes being refined; indices underlined with a solid (resp. dashed)
line represent allocated (resp. implicit) nodes.}
\label{fig:HyperTreeBinary1D-DFS-vs-Optimal-Refinements}
\end{figure}
Using short integers, of size 4 bytes (B), in order to store the
parent index child, and denoting $m$ the number of its strict nodes,
describing the topology of a hypertree thus costs at a minimum
$4(m-1)$B for the parent indices of the eldest children and $4m$B for
the indices of the eldest children themselves; this amounts to a total
cost of $8m-4$ bytes if $m\in\mathbb{N}^*$.
This theoretical minimum cost is rarely attained. The overall efficiency
of this approach is very sensitive to the topological structure of the
tree and the order in which it is traversed at construction time.
This is illustrated in
Figure~\ref{fig:HyperTreeBinary1D-DFS-vs-Optimal-Refinements}: when
the topological structure of the tree is created in DFS order, some
unnecessary allocations (namely, for nodes 4 and 5) occur; in
contrast, an optimal traversal only allocates space for strict tree
nodes. 
This worst case occurs when the last refined cell is the one which is
also the last entry during the penultimate refinement
stage, the cost for parent indices can be as high as
$4(m-1)(1+f^d)$B.

One thus obtains the following bounds for the memory cost $C(m)$,
expressed in bytes:
\[
4(2m - 1) B \le C(m) \le 4\left[1 + (m - 1)(1 + f^d) B\right].
\]
Note that the lower bound indicated above is a theoretical memory
footprint, with an ideal topology where all children of a strict node
have the same type (either all strict nodes, or all leaves) and ideal
implementation (the traversal strategy refines last all strict nodes
than only have leaf children).
Unfortunately, there is not a way to devise a traversal strategy that
is optimal for all possible topological structure of trees.

When $n$ hypertrees are embedded inside a $d$-dimensional hypertree
grid, $m\gg{n}$, this minimum cost becomes $4(2m-n)$B and is
reached when at most one common depth level is partially refined
across all hypertrees.
The maximum cost occurs in both cases when only one
hypertree is refined, and with the worst possible traversal; in this
case, the cost is $4\left[n+(m-1)(1+f^d)\right]$B for
$m\in\mathbb{N}^*$.
On the other hand, the memory footprint relative to the description of
the spatial grid, using double precision floats for the coordinates,
is least equal to $8\left(3+d\sqrt[d]{n}\right)B$ for a square or cubic
grid, and at most $8(d+n+2)$B for a linear grid.

All lower bounds theoretical mentioned above are indeed very difficult
to attain. But the lower bound defined for a topology is attain with
the ideal implementation that it is therefore the responsibility of
the developer to make this
trade-off, depending on whether additional CPU processing is
acceptable to achieve a better memory footprint, with potentially
enormous gains.
For instance, in the binary $3$-dimensional case, the memory gain factor
between this AMR description and its explicit, unstructured
all-hexahedral equivalent can range from $18$ to more than $80$.

\subsection{The Global Index Map}
A natural choice to build a concrete $\Upgamma_{d,f,E}$ is to combine
a \emph{$0$-level indexing} of the constituting hypertree roots with
the child index maps in each of these hypertrees.
For instance, the $0$-level indexing can be the lexicographic order
in~\S\ref{s:mapping-indexing}, applied to $\llbracket{E}\llbracket$.
One can then set
\[
\Upgamma_{d,f,E}(n_{\mathrm{l}};i;j;k) = n_{\mathrm{l}} + S_{i,j,k}
\]
where $S_{i,j,k}$ is the \emph{global index start} of the hypertree
object at position ${i,j,k}$ in the Cartesian grid of hypertree objects.
By construction, the restriction of $\Upgamma_{d,f,E}$ to any
particular hypertree, being piece-wise affine with unit slope, is
strictly increasing over $\mathbb{N}$ and therefore injective.
Therefore, if the $S_{i,j,k}$ are chosen so that there be
no overlap across the image spaces of these per-hypertree
restrictions, then $\Upgamma_{d,f,E}$ as a whole is injective and thus
satisfies the specification of Definition~\ref{def:global-index}.

In this setting, the global index start of each constituting hypertree
only needs to be stored as an integer offset, at the minimal
additional cost of $8$B per hypertree.
In practice, this can be achieved by constructing the hypertree grid
one hypertree object at a time, and incrementing the global index
start when moving to the next hypertree with the number of vertices in
the last constructed hypertree.

It is important to note, however, that this method to build the global
index map by means of assigning a global index start per constituting
hypertree is in no way mandatory.
Rather, our implementation provides the ability to specify an arbitrary
version of $\Upgamma_{d,f,E}$; it is the responsibility of the
developer to ensure that this map comply with the requirements of
Definition~\ref{def:global-index}.
In addition, such an explicit definition increases the total memory
footprint by the cost of representing as many integers as there are
vertices in the hypertree grid.

\subsection{The Virtual Dual}
\label{s:duality}
In order to support the widest variety of visualization algorithms, 
our first version of the hypertree grid object was implemented with a
primal/dual API, because while some visualization filters work best
processing dual grid cells, others can operate directly upon the
primal cells.
This double API thus provided visualization filters with the
alternative to traverse hypertree grids through either primal or dual
cells.

A first limitation of this design is its complexity, as two
different outside-facing data structures must 
be maintained and kept consistent.
In addition, dual grids have inherently a more complex topology than
their corresponding primal tree grids.
For instance, dual grids of $3$-dimensional tree-based AMR meshes
contain pyramids and wedges, in addition to hexahedral cells.  
Moreover, the memory footprint of the dual has proven to become
untenable when attempting to process realistically sized cases, for
the dual must be represented explicitly (in the sense of fully
described, unstructured grid), canceling all the benefits of
tree-based storage.

\begin{figure}[h]
\centering
\includegraphics[width=0.8\columnwidth]{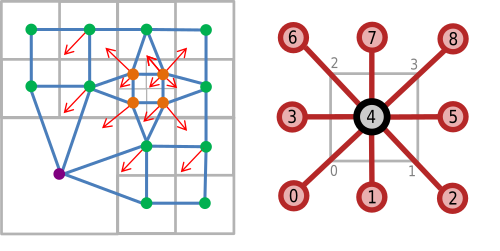}
\caption{Left: a $2$-dimensional tree-based AMR mesh $\mathcal{M}$
(gray), overlaid with $\mathcal{M}^\ast$ (blue), showing dual cell
ownership by primal vertices with orange arrows; right: cursor indices
in a $2$-dimensional Moore supercursor, used as tie-breakers for dual
cell ownership amongst the deepest primal cells.}
\label{fig:AMR2D-Dual-Ownership}
\end{figure}

\begin{algorithm}
\caption{$\mathtt{IsOwner}(s,i)$}
\label{alg:is-owner}
\begin{algorithmic}[1]                                                  
  \STATE $\delta\leftarrow\mathtt{GetDepth}(s)$
  \STATE $\omega\leftarrow\tfrac{3^d-1}{2}$
  \FORALL{$j\in\mathcal\llbracket{2^d}\llbracket$}
    \STATE $k\leftarrow \mathtt{CornerNeighborCursorsTable}[d][i][j]$
    \STATE $c\leftarrow \mathtt{GetNeighbor}(s,k)$
    \IF{$\mathtt{Masked}(c) \lor \neg\mathtt{IsLeaf}(c) \lor
         (k > \omega \land\mathtt{GetLevel}(c) = \delta)$}
      \RETURN $\mathtt{False}$
    \ENDIF
  \ENDFOR
  \RETURN $\mathtt{True}$
\end{algorithmic}
\end{algorithm}
In order to avoid cell replication in the dual grid, our method
assigns ownership of dual vertices to a single leaf, amongst the
$2^d$ that potentially touch a primal vertex in dimension~$d$:
specifically, ownership of the dual cell is assigned to the deepest of
these leaves, breaking ties in favor of the one that has the greatest
\emph{cursor index} relative to the others.
Specifically, the function that determines ownership of the dual cell
at any corner $i\in\llbracket{2^d}\llbracket$ of an arbitrary leaf
cell, at which a Moore supercursor $s$ is centered, is
explicated in Algorithm~\ref{alg:is-owner} and illustrated in
Figure~\ref{fig:AMR2D-Dual-Ownership}, left.
The $3$-dimensional integer array called
$\mathtt{CornerNeighborsCursorTable}$ is a table that provides, given a
$d$-dimension Moore supercursor and a corner index, the indices of the
cursors that surround said corner.
that surround  centered at a cell is a corner-to-leaf traversal  table
to retrieve the $2^d$ indices of all the cell cleaves touching a given
corner of a given cell.
For example in dimension~$2$, as illustrated in
Figure~\ref{fig:AMR2D-Dual-Ownership}, right,
\begin{align*}
\mathtt{CornerNeighborCursorsTable}[2][0] & = \{ 0; 1; 3; 4\} \\
\mathtt{CornerNeighborCursorsTable}[2][1] & = \{ 1; 2; 4; 5\} \\
\mathtt{CornerNeighborCursorsTable}[2][2] & = \{ 3; 4; 6; 7\} \\
\mathtt{CornerNeighborCursorsTable}[2][3] & = \{ 4; 5; 7; 8\}.
\end{align*}
Although this method keeps the additional memory footprint at
the strict minimum, by avoiding dual cell duplication, it
is not sufficient to prevent memory overruns even with relatively
modestly-sized AMR meshes as a result of the unstructured nature of
the dual mesh.
As a result, we completely revised our initial approach, by
retaining the main idea of utilizing duality as a natural means to
process conforming cells when necessary for the considered
visualization technique, while adding the two following design
requirements:
\begin{enumerate}[(i)]
\item
ready access to individual dual cells when required,
\item
storage of the entire dual mesh is prohibited.
\end{enumerate}
Our new methodology thus consists of utilizing a \emph{virtual dual},
of which only one cell can be stored at any point in time.
Provided an efficient way to generate, on demand, such individual cells
from the virtual dual can be devised, then all memory footprint
problems will vanish.
Meanwhile, and by the same token, it will remain possible to apply
visualization techniques that must, by design, operate on the cells of
a conforming mesh.
In this goal, we retained from our earlier approach the notion of
dual item ownership, with the subtle yet important difference that it
is expressed in terms of primal cell (and hence dual vertex)
ownership of dual cells.
This trade-off comes obviously at the price of added computational
cost for the benefit of memory footprint, as the dual is not computed and
stored once and for all.

\subsection{A Hierarchical Approach to Cursors}
\label{s:hierarchy-cursors}
In our first attempt at using cursors designed to traverse hyper tree
grid objects, we sought to handle all cases at once by implementing 
a supercursor, designed as a $3^3$ grid of cursors, with the center
cursor simply performing a DFS traversal while tracking all possible
26 adjacent nodes (some of which remaining empty depending on the
values of $f$ and $d$).
In order to achieve this vicinity tracking, the methodology was making
use of pre-computed look-up tables able to tell, for each child being
visited, how to populate the new grid of cursors (using the same
disambiguation rules as described in~\S\ref{s:cursors-supercursors}).
Such a supercursor can indeed by initialized at the root level of any
given hypertree, by considering the placement of the corresponding
root cell within its $d$-dimensional embedding in the Cartesian grid
of all root nodes.

\begin{figure}[h]
\centering
\includegraphics[width=0.95\columnwidth]{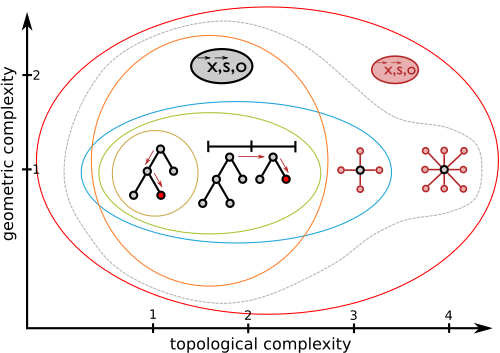}
\caption{Venn diagram describing the possible subsets of topological
(tree DFS, tree grid DFS, von Neumann and Moore neighborhoods)
and geometric (none vs. cell coordinates and sizes) features that grid
cursors and supercursors can have.
Currently implemented, and used cursors are:
\texttt{\textcolor{brown}{HyperTreeCursor}},
\texttt{\textcolor{olive}{HyperTreeGridCursor}},
\texttt{\textcolor{orange}{GeometricCursor}},
\texttt{\textcolor{cyan}{VonNeumannSuperCursor}}, and
\texttt{\textcolor{red}{MooreSuperCursor}}.
The earlier ``simple'' supercursor is depicted in dashed gray.}
\label{fig:HyperTreeCursors}
\end{figure}

This early implementation demonstrated the theoretical soundness of
the approach, as our proof-of-concept dual mesh algorithm
demonstrated~\cite{carrard:imr21}.
However, it quickly became obvious that maintaining a full
neighborhood of cursors in all cases, containing all possible
topological and geometric information, was computationally too costly.
At the same time, we gradually came to realize that not all filters
needed all this wealth of information.
We therefore set about distinguishing between the different natures,
geometric and topological, and the various degrees of information that
hypertree grid cursors and supercursors could possibly provide.
The results of this effort are summarized from a qualitative vantage
point in Figure~\ref{fig:HyperTreeCursors}, where the 
horizontal axis distinguishes, left to right, between 4 increasing
levels of complex topological information; meanwhile, the vertical
axis separates, from bottom to top, between the two different levels
of geometric information that could conceivably be needed by hypertree
grid filters.

Specifically, we have the following levels of topological information,
from least to most complex:
\begin{enumerate}
\item
The simplest type of hypertree traversal we can conceive of has
DFS type, where child/parent connectivity is all that is needed to
traverse the entire tree.
\item
Because the main object for our stated purposes is not the hypertree
\emph{per se}, but the hypertree grid, one level of topological
information that can naturally be added atop the previous one is the
ability to traverse horizontally between hypertree root cells within
the grid thereof\footnote{note that this case still amounts to DFS
traversal, by conceiving of a meta-root above all
actual tree roots.}.
\item
As explained in~\S\ref{s:cursors-supercursors} some filters need to
know the Von Neumann neighborhood of any given cell in order to
process it.
Therefore, the ability to keep track of such neighborhoods while
traversing the hypertree grid is the next level of topological complexity.
\item
Finally, as has also been discussed in~\S\ref{s:cursors-supercursors},
all filters relying on dual cell construction require knowledge of
Moore neighborhoods.
\end{enumerate}

Meanwhile, our geometric complexity stack is much simpler, for it only
distinguishes between two different cases, as illustrated along the
vertical axis of Figure~\ref{fig:HyperTreeCursors}, as follows:
\begin{enumerate}
\item
In its simplest form, geometric information is empty; in other words,
the considered traversal does not need access to any of the geometric
features, resulting from the $3$-dimensional embedding of the
currently traversed hypertree object.
This happens most notably while constructing, on-the-fly, the
tree-structure of a hypertree grid output while traversing a hypertree
grid input whose geometric information is the only that which is
needed.
\item
The only other type of geometric information, consistent with our
design choices explained in~\S\ref{s:hypertree}, is the geometric
embedding triple
$(\overrightarrow{x};\overrightarrow{s};o)\in(\mathbb{R}^3)^2\times\mathbb{N}$.
\end{enumerate}
Note, however, that this rather simple geometric information stack
could be enriched as needed, should we decide to support more complex
geometric information such as, for instance, the
$(\overrightarrow{x};\overrightarrow{s};\overrightarrow{v})\in(\mathbb{R}^3)^3$,
also considered in~\S\ref{s:hypertree}, but currently not supported.
Similarly, should other types of connectivities arise in the future,
these would readily find their place along the topological complexity scale.

These two complexity stacks can be viewed as independent in the
context of tree cursors.
We can then make the convention to lay them out along two orthogonal
axes, and to represent their combinations of interest as Venn diagrams,
with the additional convention that a more complex cursor always
contains all features of the less complex ones.
We thus obtain the conceptual $2$-dimensional representation of
Figure~\ref{fig:HyperTreeCursors}, where any given cursor, or
super-cursor, can be represented in terms of a Venn diagram
containing the needed features, both geometric and topological.
Indeed, this schematic outlines the corresponding properties of
the five cursors and supercursors we came to realize were necessary for
our considered applications thus far, as follows:
\begin{description}
\item[\texttt{\textcolor{brown}{HyperTreeCursor}}:]
hypertree traversal with DFS, without geometric information.
\item[\texttt{\textcolor{olive}{HyperTreeGridCursor}}:]
hypertree grid traversal with DFS, without geometric information.
\item[\texttt{\textcolor{orange}{GeometricCursor}}:]
hypertree grid traversal with DFS, with geometric information at
cursor center.
\item[\texttt{\textcolor{cyan}{VonNeumannSuperCursor}}:]
hypertree grid traversal with both DFS and von Neumann connectivity,
with geometric information at supercursor center.
\item[\texttt{\textcolor{red}{MooreSuperCursor}}:]
hypertree grid traversal with both DFS and Moore connectivity,
and geometric information at supercursor center and for all vicinity cursors.
\end{description}
Note that our earlier, one-size-fits-all, ``simple'' supercursor is
somewhere between the two latter ones, providing hypertree grid traversal
with both DFS and Moore connectivity, but with geometric information
only at the center of the supercursor.
This is because the ownership of dual cells by explicitly stored dual
points instead of primal cells, as explained
in~\S\ref{s:duality}, eliminates the need to retrieve
neighbor cell geometric information when generating a dual cell on the
fly.

\begin{figure}[h]
\centering
\includegraphics[width=0.99\columnwidth]{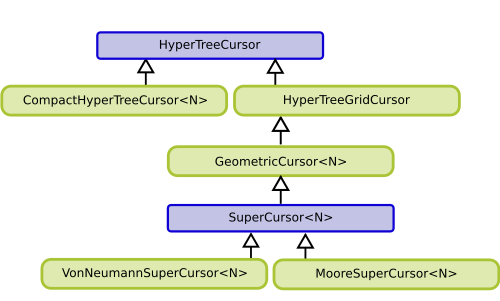}
\caption{Inheritance diagram of the hierarchy of cursors and
supercursors implementation; classes shown in green (resp. blue) are
concrete (resp. abstract).}
\label{fig:inheritance}
\end{figure}
The granularity offered by our novel hierarchy of cursors and
supercursors not only allows for the fine-tuning of algorithms in
order to optimize execution speed, but can also be extended to include
other combinations of properties, or even new properties, as target
applications will command.
This hierarchy is implemented following the inheritance diagram
shown in Figure~\ref{fig:inheritance}. 

\subsection{Supercursor Traversals}
As explained in ~\S\ref{s:cursors-supercursors}, a hypertree
cursor implementation should implement the \texttt{ToParent()} and
\texttt{ToChild(i)} operators in order to allow for movement up and
down a tree.
Meanwhile, when it is only needed to visit all tree vertices in
DFS order, it suffices to be able to position the cursor at the
root of every hypertree, and then recursively call \texttt{ToChild(i)}
to perform the traversal.
This is our primary mode of operation, which \emph{de facto}
eliminates the need for a \texttt{ToParent()} operator, which is thus
replaced in practice with a \texttt{ToRoot()} function to position the
cursor at the root of each constituting hypertree.

In the case of cursors that are not supercursors
(cf.~\S\ref{fig:inheritance}), both methods are relatively easy to
conceive of, and are therefore not discussed in detail here.
Furthermore, \texttt{ToRoot()} is also rather simple to implement
for supercursors, given the Cartesian layout of a hypertree grid at
the level of the roots.
The matter is more complicated for \texttt{ToChild(i)}, however,
because all neighborhood cursors must be updated upon descent of the
supercursor into a child node.
This update cannot be done \emph{a priori}, because neighborhoods no
longer have a Cartesian grid structure as soon as depth is non-zero;
Instead, the neighborhood of a child must be explicitly computed from
that of its parent.
This task may seem daunting at first glance, but we devised an
approach based on pre-computed \emph{traversal tables} that greatly
facilitates these updates.

Given a supercursor $s$ pointing at a cell $C$, each of the children
of this cell are uniquely identified by their respective child index
$i\in\llbracket{f^d}\llbracket$, as explained in
Definition~\ref{def:child-index}.
Now, given $f$ and $d$, there exists a unique mapping from the entries
of the supercursor of child $C_i$ into those of its parent $C$.

\begin{figure}[ht!]
\centering
\includegraphics[width=0.65\columnwidth]{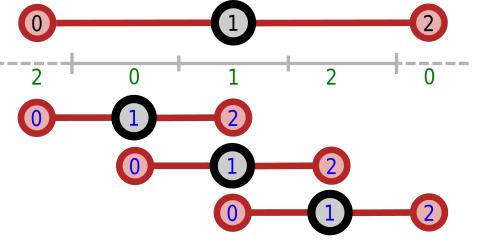}
\caption{Supercursor parent/child and child/child relationships when
$d=1$ and $f=3$: cursor indices in black (parent) or blue (children), child indices in
green.}
\label{fig:SuperCursorsParentAndChildren-d1-f3}
\end{figure}
Consider for instance the easiest $1$-dimensional case, where there is
no difference between Von Neumann and Moore supercursors, each of
these containing $3^1=2\times1+1=3$ cursors.
Figure~\ref{fig:SuperCursorsParentAndChildren-d1-f3} illustrates
this case, with a solid gray line representing a coarse cell~$C$
divided with $3$ children cells; child indices are indicated in
green.
Potential neighbor cells are shown on both sides with dashed lines;
child indices adjacent to the cell of interest are labeled as well.
In the same figure are also pictured the supercursors centered at~$C$
(above the line) and those centered at each of its children (below).
In this case, the cursor with index 0 (i.e., pointing to the left) of
the supercursor $s_0$ centered at child $C_0$ will point towards to either
the same cell as cursor with index~$0$ of the supercursor $s$ centered
at parent cell~$C$, or to one of its children.
Meanwhile, the two other cursors of $s_0$ will point to either the
same cell as cursor with index~$1$ of~$s$, or to one of its children.
This logic thus yields the following map, between child and parent
cursors, for child~$0$: $0\mapsto0$, $1\mapsto1$, and $2\mapsto1$,
which we denote $(0;1;1)$ in compact form.
One can easily deduce the corresponding maps for the children of~$C$
with indices~$1$ and~$2$, by reading the blue indices of
Figure~\ref{fig:SuperCursorsParentAndChildren-d1-f3} from left to
right, mapping them to the cursor indices in black for the
corresponding parent supercursor. 
When concatenated in child index order, these $3$ maps provide the
\emph{child cursor to parent cursor table} for the case
where $d=1$ and $f=3$, i.e. $(0;1;1;1;1;1;1;1;2)$.

It is important to note than the \emph{or to one of its children}
clause above may occur only when the cell to which a cursor $c$ of the
supercursor is pointing is not a leaf.
As explained in~\S\ref{s:cursors-supercursors}, said cursor $c$ cannot
point to a cell of depth greater than that of~$C$, but a supercursor
centered at child of~$C$ can point to cell exactly at most one level
deeper.
When this situation occurs, $c$ must be descended into the adequate
child of the parent cell neighbor in order to retrieve the
corresponding vicinity cursor of the child supercursor.
For example, in Figure~\ref{fig:SuperCursorsParentAndChildren-d1-f3},
if cell $C_-$ to which cursor with index~$0$ in the supercursor
of~$C$ is pointing is not a leaf, then cursor with index~$0$ in the
supercursor of child cell~$C_0$ must point to child with
index~$2$ of $C_-$.
Another type of map is therefore required to perform the descent into
the relevant children of coarse cells whenever necessary.
In the current example, $C$ is coarse, hence cursor with index~$1$ in
$s_0$, (i.e., the center cursor) will point to $C_0$ itself, i.e. to
the child cell with index~$0$.
Similarly, $C$ being coarse, cursor~$2$ of $s_0$ will point at child
with index~$1$ of~$C$.
We hence obtain the following map in compact form: $(2;0;1)$.
The corresponding maps for children $C_1$ and $C_2$ are obtained
accordingly, reading the green indices of  
Figure~\ref{fig:SuperCursorsParentAndChildren-d1-f3} from left to right,
mapping them to the child cursor indices in blue for the corresponding
child supercursor. 
When concatenated in child index order, these $3$ maps provide the
\emph{child cursor to child index table} for the case where $d=1$ and
$f=3$, i.e. $(2;0;1;0;1;2;1;2;0)$.
 
\begin{figure}[ht!]
\centering
\includegraphics[width=0.49\columnwidth]{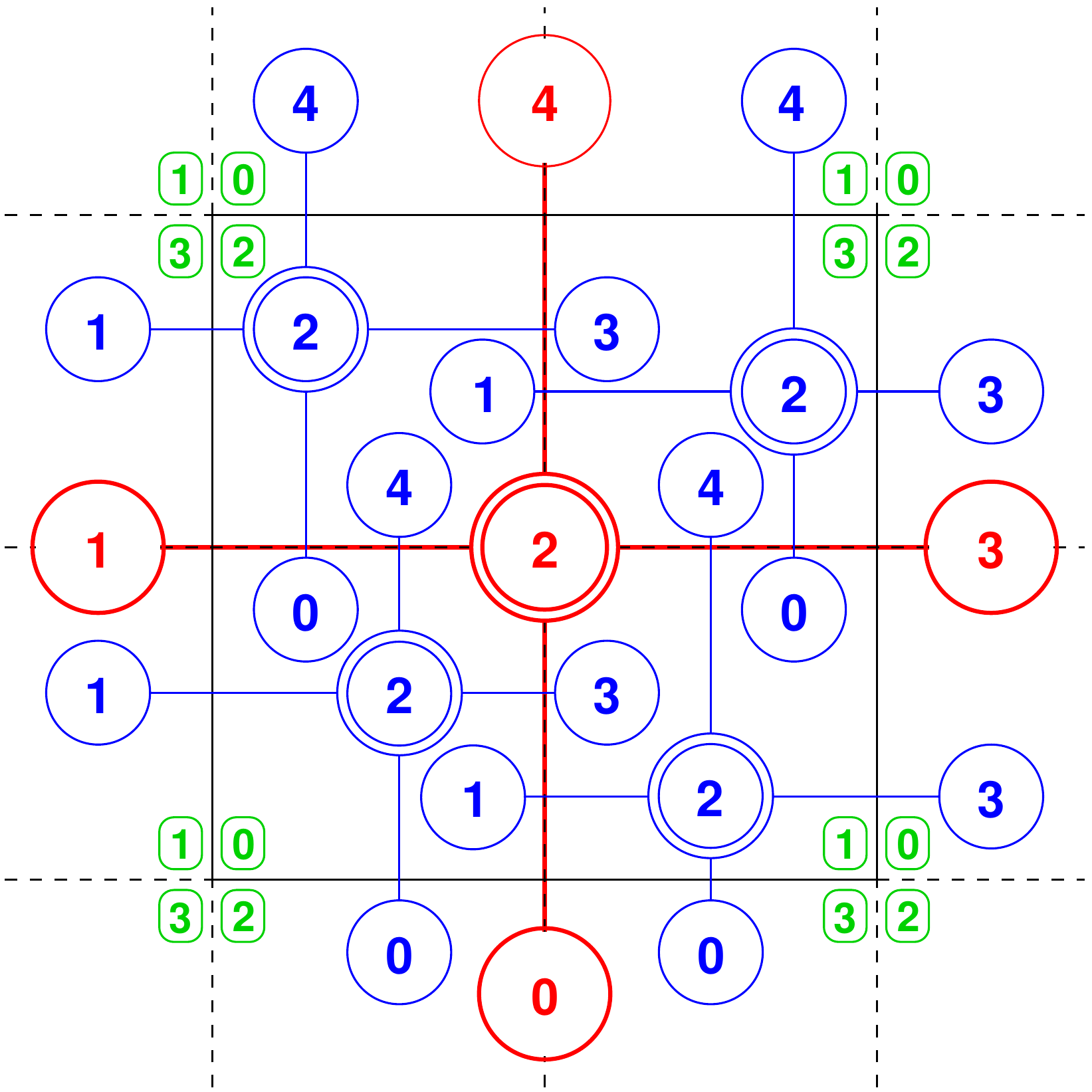}
\includegraphics[width=0.49\columnwidth]{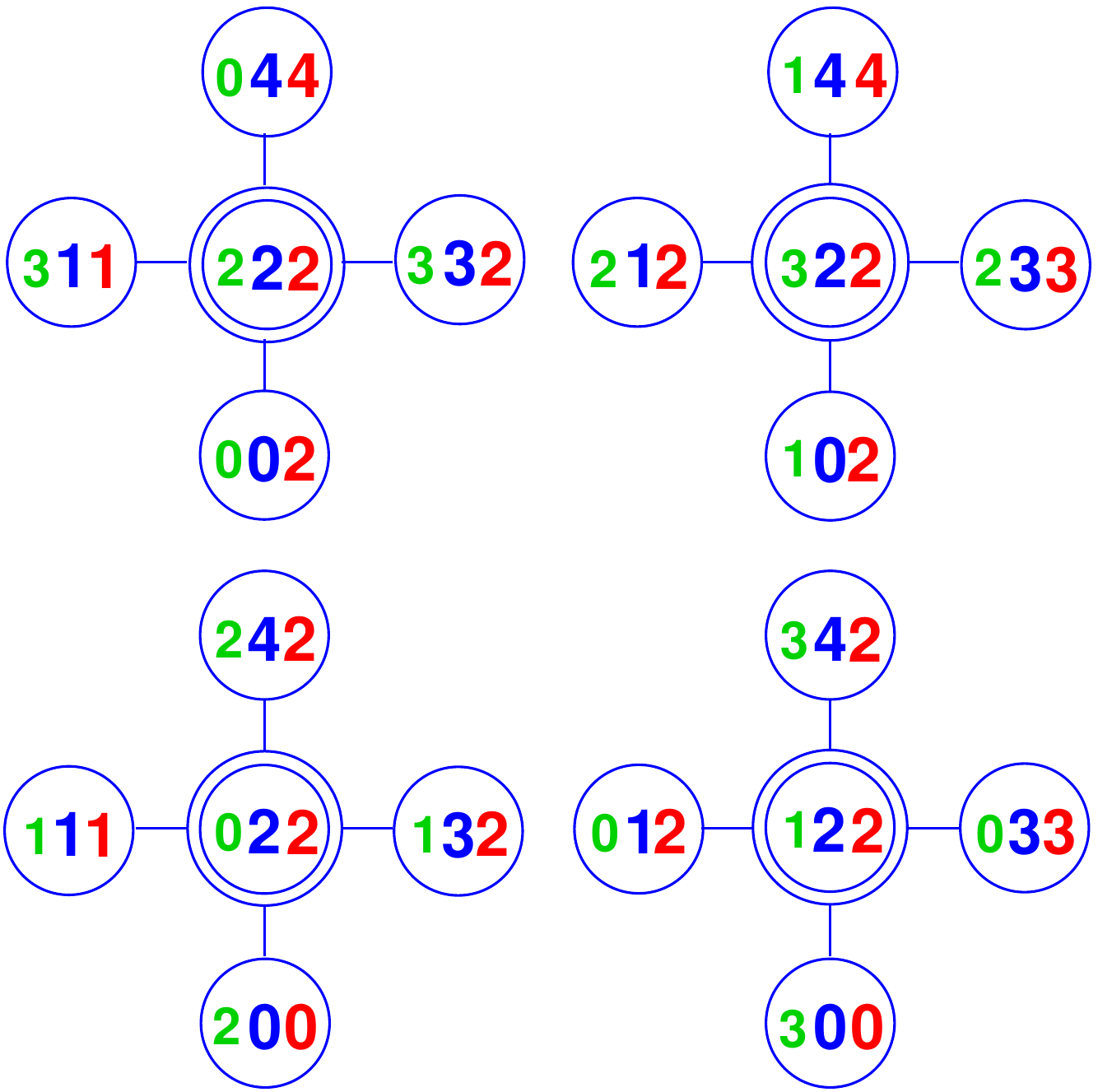}
\caption{Von Neumann supercursor $d=2$ and $f=2$: parent/child and
child/child relationships (left), corresponding traversal table
(right) with vicinity cursor indices in blue, child indices in
green, and parent cursor indices in red.}
\label{fig:VonNeumannSuperCursor-d2-f2}
\end{figure}
In order to entertain the reader, we provide the corresponding
diagrams for the Von Neumann supercursor when $d=2$ and $f=2$ in
Figure~\ref{fig:VonNeumannSuperCursor-d2-f2}.
Such schematics can be used to derive all traversal tables, for all
types of supercursors and all possible values of $d$ and $f$.
Drawing all possible cases would however be a tedious as well as
error-prone task, so we implemented a Python script
in order to generate the $2804$ entries filling the $24$
($2\times2\times2\times3$) possible tables.
Traversal table initialization can thus be performed only once at
supercursor construction time, based on template parameter value and
type of supercursor.
This methodology thus ensures code correctness, as well as optimal
execution speed for table entry retrieval is only a matter of random
access in a small static arrays.

When endowed with these pre-computed tables, updating supercursors
when performing a DFS traversal becomes easy: given a supercursor~$s$
centered at a given coarse cell, all of its cursors are copied in
temporary storage to avoid memory stomping as cross-permutations will
occur.
Then, given a child index~$i$, for each cursor index~$j$ the
corresponding cursor index~$k$ in the parent supercursor is retrieved
from the child cursor to parent cursor table.
The cursor with index~$k$ of the parent supercursor, previously copied
as~$c[k]$, is assigned to the child cursor and if~$c[k]$ points at a
leaf, then the update is complete. 
However, if~$c[k]$ points to a coarse cell, then it must be
descended into, using the appropriate child index, retrieved from the
child cursor to child index table.

\begin{algorithm}
\caption{$\mathtt{SuperCursorToChild}(s,i)$}
\label{alg:to-child}
\begin{algorithmic}[1]
  \STATE $n\leftarrow\mathtt{GetNumberOfCursors}(s)$
  \FORALL{$j\in\mathcal\llbracket{n}\llbracket$}
    \STATE $c[j]\leftarrow\mathtt{GetCursor}(s,j)$
  \ENDFOR
  \STATE $C\leftarrow\mathtt{GetChildCursorToChildTable}(s,i)$
  \STATE $P\leftarrow\mathtt{GetChildCursorToParentCursorTable}(s,i)$
  \FORALL{$j\in\mathcal\llbracket{n}\llbracket$}
    \STATE $k\leftarrow P[j]$
    \STATE $\mathtt{GetCursor}(s,j)\leftarrow c[k]$
    \IF{$\neg\mathtt{IsLeaf}(c[k])$}
      \STATE $\mathtt{CursorToChild}(\mathtt{GetCursor}(s,j),C[j])$
    \ENDIF
  \ENDFOR
\end{algorithmic}
\end{algorithm}
This scheme is summarized in Algorithm~\ref{alg:to-child}. 
We explicitly distinguish between the
$\mathtt{SuperCursorToChild}(c,i)$ and $\mathtt{CursorToChild}(s,i)$
methods in order to emphasize that this method is \emph{not}
recursive: when descent into a child must performed, it is only
performed on a cursor of the supercursor, not on the supercursor
itself.
Note that this formulation of the algorithm ignores,
for the sake of legibility, everything that regards the geometric
updates which must also be performed.

\subsection{Filters}
\label{s:filters}
We now discuss our methodology to \emph{filtering}, i.e. applying
visualization and data analysis algorithms, to hypertree grid objects.
We begin with the case of geometric transformations, which can be
especially efficiently addressed thanks to the notion of geometric
embedding.
We then explain our two-pass approach, based on a pre-selection stage,
used to improve execution speed for those algorithms that rely on
heavyweight supercursors.
This section closes with a high-level description of the currently
implemented filters, whose choice was dictated by actual
analysis needs rather than for the sake of academic interest, and how
they relate to the previously discussed cursors and supercursors.
\subsubsection{Geometric Transformations}
\label{s:gemometric-transformations}
We recall that, as defined in~\S\ref{s:hypertree}, we can represent
the geometry of any arbitrary rectilinear, tree-based AMR by means of
the $3$-dimensional embedding
$(\overrightarrow{x};\overrightarrow{s};o)\in(\mathbb{R}^3)^2\times\mathbb{N}$
of its hypertree grid equivalent.
Because we restricted ourselves to the case of axis-aligned
geometries, not all geometric transformations can be represented with
this model: for example, a projective transformation will not
transform, in general, a rectilinear hypertree grid into another.
In fact, not even all affine transformations are suitable: as a
result of our choice to only support axis-aligned grids, arbitrary
rotations cannot be supported within our current framework either.
Nonetheless, restricting possible transformations to that preserve
alignment with the coordinate axes entails no loss of generality
because, the AMR grids we aim to support are assumed to be
axis-aligned by design (cf.~\S\ref{s:hypertree}).

For example, it is easy to see that all axis-aligned reflections, i.e.,
symmetries across a hyperplane that is normal to one coordinate axis,
comply with the requirements above, being affine and preserving
parallelism with all coordinate axes.
We call \texttt{AxisReflection} such a transformation filter in our
nomenclature.
Furthermore, the reflection across a hyperplane in dimension~$d\le3$, that
is normal to axis $i\in\llbracket0;d\llbracket$ and has coordinate
$\omega\in\mathbb{R}$ can be embedded in dimension~$3$ as follows:
\[
\begin{array}{rccc}
r_{i,\omega}: 
& \mathbb{R}^3  & \longrightarrow & \mathbb{R}^3 \\
& (x_0;x_1;x_2) & \longmapsto & (x_0';x_1';x_2')
\end{array}
\]
where
\[
\forall k\in\{0;1;2\} \quad
\renewcommand{\arraystretch}{1.2}
\left\{\begin{array}{ll}
k  =  i & \Rightarrow x_k' = 2\omega - x_k, \\
k\neq i & \Rightarrow x_k' = x_k.
\end{array}\right.
\]
It thus follows that the image by $r_{i,\omega}$ of the
$3$-dimension geometric embedding of an hypertree object is
\[
r_{i,\omega}(\overrightarrow{x};\overrightarrow{s};o)
= 
(r_{i,\omega}(\overrightarrow{x});\overrightarrow{s}_{-i};o),
\]
where $\overrightarrow{s}_{-i}$ denotes the vector equal to
$\overrightarrow{s}$, save for its $i$-th coordinate which is opposed
to that of $\overrightarrow{s}$.
Therefore, the geometric embedding of the image by $r_{i,\omega}$ of
an hypertree grid is exactly the collection of all image geometric
embeddings of its constituting hypertrees.
Axis-aligned reflection of hypertree grid objects can thus be
implemented in a way that only operates upon the geometric embeddings
using the very simple formula above.
As a result, such an implementation is both extremely fast and memory
efficient, for all it needs to do is create a new
array of transformed coordinates along a single axis for the
geometric embeddings of its constituting hypertrees.
Meanwhile, the topological structures of said hypertrees only have to
be shallowly copied.

\subsubsection{Dual-Based Filters}
\label{two-stage-filters}
As explained in~\ref{s:duality}, we devised the concept of virtual
dual, in order to extend the the range of applicability of our
original dual-based approach to include large-scale meshes.
The elements of this virtual dual are thus to be generated, processed
and discarded at once as the filter traverses the input grid.
In order to generate the dual cell associated with an arbitrary primal
vertex (\emph{corner}) as illustrated in
Figure~\ref{fig:AMR2D-Dual-Ownership}, left, a filter must be able to
iterate over all primal cells sharing that corner.

Traversal of the input AMR mesh is performed over the
vertices of the corresponding hypertree grid using cursor objects
discussed in~\ref{s:cursors-supercursors}.
Therefore, on-the-fly dual cell creation occurs by iterating over the
all corners of all input primal cells and, for each such corner,
iterating over all primal cells having it as a corner.
In dimension~$2$ for instance, there can be 2 across-edge neighbors
and 1 across-corner neighbor to a primal cell that share a given
corner thereof.
In dimension~$3$, 3 across-face neighbors can also exist, as well an
one additional across-edge neighbor.
The cursor must thus provide Moore neighborhoods so that all those
types of neighbors of a cell are made available when iterating around
one of its corners.

In addition, when a dual cell must actually be generated, based on the
ownership rules introduced in~\ref{s:duality}, its vertices are, by
definition, located at primal cell centers (and possibly moved the
primal boundary when dual adjustment is be performed).
As a result, the cursor must also provide access to the geometric
information of all neighbors.
Both features, topological and geometric, are provided by the Moore
supercursor which is thus required by all dual-based filters.
This super-cursor is the most complex in our hierarchy of cursors, and
every traversal operation onto it requires many operations, with a
computational cost that becomes quickly prohibitive as input mesh size
increases.
This can result in losses in interactivity detrimental to the
analysis process, or even in unacceptable execution times.

\begin{figure}[htb]
\centering
\begin{minipage}[t]{0.45\columnwidth}
\centering
\vspace{0pt}
\includegraphics[width=0.99\textwidth]{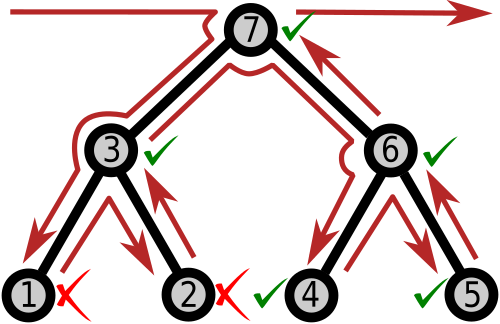}
\end{minipage}
\hfil
\begin{minipage}[t]{0.45\columnwidth}
\centering
\vspace{0pt}
\includegraphics[width=0.99\textwidth]{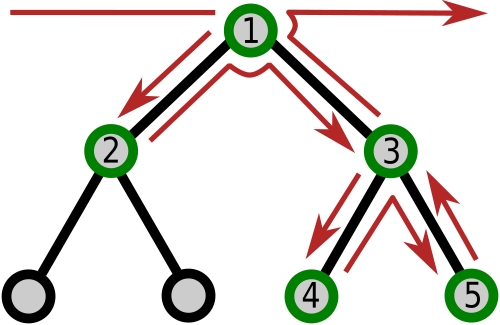}
\end{minipage}
\caption{Stages of a two-stage filter applied to one constituting
hypertree within a binary hypertree grid. 
Left: pre-processing stage with post-order DFS traversal, using a
lightweight cursor, selecting vertices check-marked in green.
Right: main stage with pre-order DFS traversal with a heavier
cursor, only across pre-selected vertices.
The indices reflect the order in which vertices are processed by each
stage.}
\label{fig:TwoStageFilter}
\end{figure}
In order to circumvent this difficulty, we devised a two-pass approach
where a more lightweight cursor is used to traverse the entire mesh in
a pre-processing stage, selecting only those cells that are concerned
by the algorithm.
The dual-based computation is thus only performed in the subsequent
processing stage, where only those pre-selected parts of the grid are
actually traversed by the most expensive Moore supercursor.
Specifically and as illustrated in Figure~\ref{fig:TwoStageFilter}, 
the pre-selection stage uses \emph{post-order} DFS traversal, in order
to propagate upwards per-branch selection (and possibly aggregated
attribute information as well), whereas the main stage is
performed with \emph{pre-order} DFS fashion, immediately processing
the pre-selected cells in the order in which they are reached when
skipping non-selected branches.
Albeit more complex in appearance, this two-stage approach can in fact be
dramatically more efficient than a direct traversal of the input grid
with the most complex cursor, provided a clever pre-selection
criterion not requiring neighborhood information be contrived.
The key success factor to this approach thus rests on devising a
criterion that is easy to compute with minimal information and yet is
discriminatory enough so as to avoid as many false positives as
possible (while false negatives will result in an incomplete output).

\subsubsection{Concrete Filters}
\label{s:concrete-filters}
We now provide a brief overview of the filters we have developed so
far, as concrete instances of our cursor-based general methodology.
This list can, and most likely will, be extended as dictated by
tree-based AMR post-processing needs.
\begin{description}
\item[\texttt{AxisCut}:]
produce a $2$-dimensional hypertree grid output from a $3$-dimensional
input, comprising the intersection of all cells in the latter that are
intercepted by an axis-aligned plane.
The output has an associated material mask only when the input has
one.
\item[\texttt{AxisClip}:]
clip, i.e., mask out all input cells that do not fulfill a
geometric condition that can take three forms: hyperplane, rectangular
prism (shorthand \emph{box}), or quadratic function.
In hyperplane mode, only those leaf cells that are either
intersected by said hyperplane or wholly within a prescribed
half-space that it defines are retained.
A similar selection process occurs in box mode, based on whether
cells are intersected by said box or located entirely in its interior.
In quadratic mode, a leaf cell is retained if and only if
said function takes on positive values at all corners of this cell.
The hypertree grid output always has a material mask
even when the input does not.
\item[\texttt{AxisReflection}:]
already presented as a geometric transformation filter exemplar
in~\S\ref{s:gemometric-transformations}.
\item[\texttt{CellCenters}:]
generate the set of points consisting of the centers of the leaf cells
in a hypertree grid, with the option to make it also a polygonal data
set containing only vertex elements.
\item[\texttt{Contour}:]
compute polygonal data sets representing iso-contours corresponding to
a set of given values for the cell-wise attribute, using a dual-based
approach with a pre-selection criterion discussed in
detail in~\S\ref{s:isocontouring}.
\item[\texttt{DepthLimiter}:]
stop the descent into each of the constituting hypertrees whenever
either a leaf or the requested maximum depth are 
reached; in the latter case, a leaf is issued to replace the reached
node, and therefore all its descendants too.
The output is a hypertree grid that has a material mask only if the
input does as well.
\item[\texttt{Dual}:]
generate the entire dual mesh, possibly adjusted.
For the reasons developed in~\S\ref{s:dual}, this filter should never
be used with sizable hypertree grid inputs, but only for prototyping
or illustration purposes.
\item[\texttt{Geometry}:]
generate the outside surface of a hypertree grid as a polygonal data
set, in particular for rendering purposes.
Note that, already memory costly in dimension~$3$, this conversion
into an unstructured mesh can, in dimension~$2$, create an output whose
footprint is orders of magnitude larger than that of the hypertree
grid input.
\item[\texttt{PlaneCutter}:]
similar to the \texttt{AxisCut}, except that it can take an arbitrary
plane as cut function, to produce a polygonal data set output.
This filter has two modes of operation: primal or dual.
In primal mode, both topology and geometry of the original leaf cells are
preserved, hereby ensuring that no interpolation error
occur and that the cut planes extend to the primal boundary, at the
topological cost of producing T-junctions wherever the cut plane
intercepts an interface between cells at different depths.
\item[\texttt{Threshold}:]
produce a hypertree grid output with an associated material mask, even
when the input does not have any, in order to mark out all cells whose
attribute value is not within a specified range.
\item[\texttt{ToUnstructured}:]
generate a fully explicit unstructured grid data set whose elements
are exactly the leaf cells of the input hypertree grid, represented as
rectangular prisms (i.e., lines, \emph{quads}, or \emph{voxels}
depending on the dimensionality of the input).
The output thus has exactly the same geometric support as the input;
it is not a conforming mesh due to the presence of T-junctions.
It is also prohibitively expensive for sizable AMR meshes and shall
thus only be used for prototyping or illustration purposes.
\end{description}

\begin{table*}[!h]
  \renewcommand{\arraystretch}{1.2}
  \caption{Cursors vs. filters: unless otherwise mentioned, check-marks
correspond to the cursors used to iterate over the input grid, when
needed (which is not always the case). $d$ denotes the dimensionality
of the input grid.
Cursors are arranged left to right in increasing order of complexity.
$^\ast$Note that for the \texttt{Dual} filter, old ``simple''
super-cursor was sufficient because no neighbor geometry information
is necessary as dual points are all computed and stored explicitly.}
  \label{t:available-operations}
  \centering
    \begin{tabular}{@{}l@{}ccccc@{}}
    \hline \\[-14pt]
         & \texttt{TreeCursor}
         & \texttt{TreeGridCursor}
         & \texttt{GeometricCursor}
         & \texttt{VonNeumannSuperCursor}
         & \texttt{MooreSuperCursor} \\
    \hline
    \hline
    \texttt{AxisClip}           & \checkmark (output) & & \checkmark \\\hline
    \texttt{AxisCut}            & \checkmark (output) & & \checkmark \\\hline
    \texttt{AxisReflection}     & & & \\\hline
    \texttt{CellCenters}        & & & \checkmark \\\hline
    \texttt{Contour}            & & \checkmark (pre-processing) & & & \checkmark \\\hline
    \texttt{DepthLimiter}       & \checkmark (output) & \checkmark \\\hline
    \texttt{Dual}               & & & & & \checkmark$^\ast$ \\\hline
    \texttt{Geometry}           & & & \checkmark ($d<3$) & \checkmark ($d=3$) \\\hline
    \texttt{PlaneCutter} \begin{tabular}{@{}l@{}}primal\\dual\end{tabular}
         & & \begin{tabular}{@{}c@{}}\\\checkmark(pre-processing)\end{tabular} 
         & \begin{tabular}{@{}c@{}}\checkmark\\\vphantom{a}\end{tabular} 
         & & \begin{tabular}{@{}c@{}}\\\checkmark\end{tabular} \\\hline
    \texttt{Threshold}          & \checkmark (output) & \checkmark \\\hline
    \texttt{ToUnstructuredGrid} & & & \checkmark \\\hline
    \end{tabular}
\end{table*}
These filters are implemented using their respective minimal cursors
within the set described in~\S\ref{s:cursors-supercursors}.
The correspondence between filters and cursors is provided in
Table~\ref{t:available-operations}; it is left to the reader to
examine why these relationships are indeed both correct and minimal.

\subsection{Iso-Contouring}
\label{s:isocontouring}
We conclude this methodological discussion by emphasizing the case of
iso-contouring, because it is arguably one of the most widely used
amongst all existing visualization techniques, while being especially
difficult to perform on AMR grids -- in practice, impossible when
dealing with large grids if they must be converted to an explicit grid
prior to iso-contouring.
It is important to mention that we iso-contour hypertree grid
attribute fields by considering only their values at leaf nodes.
This design choice is made in order to simplify a complex
problem.
Note however that subsequent implementations could be allowed 
to take into account field values at strict tree nodes as well.
That said, there is no known, efficient iso-contouring algorithm for general
polyhedral meshes.
Instead, the canonical approach is to subdivide polyhedra into
simplices which are subsequently iso-contoured\footnote{provided
the interpolation scheme be linear, an axiom which we make for the
type of elements we want to support, and which therefore we will not
discuss further here.}.
As discussed in~\S\ref{s:foundations}, this approach is prohibitive in
terms of memory footprint and execution time.
It is therefore natural to consider a dual-based approach to tackle
iso-contouring of hypertree grids.
Therefore, as explained in~\S\ref{two-stage-filters}, the
computational efficiency of the algorithm rests upon an astute
pre-selection criterion to decide whether a cell may be
intercepted by an iso-contour without any information retrieval
concerning its neighbors.

Given a hypertree grid $\mathcal{H}$ with $n_v$ vertices and an array
$C$ of iso-values, our selection criterion defines $\lvert{C}\lvert$
Boolean arrays with length $n_v$ called \emph{sign arrays}.
For every value in $C$ with index
$j\in\llbracket\lvert{C}\lvert\llbracket$, the  
corresponding signed array is denoted $S_j$.
The goal of each $S_i$ is to capture the relative position of the
field of interest at all tree vertices, with \texttt{True}
(resp. \texttt{False}) when the cell-centered\footnote{or, for the
sake of iso-contouring, considered as such.} value is greater
(resp. smaller) than $C[j]$.
We also define another Boolean array, $T$, called the \emph{truth
array}, with length $n_v$ has well.
$T$ is global to the entire set of iso-contours and is used to
pre-select tree cells that  will be immediately iso-contoured by the
main processing phase. 
Only one such~$T$ is used across all iso-contours because a dual cell
must be generated when required by at least one iso-value. 

\begin{algorithm}[ht!]
\caption{$\mathtt{RecursivelyPreProcessTree}(c)$}
\label{alg:recursively-pre-process} 
\begin{algorithmic}[1]
  \STATE $i\leftarrow\mathtt{GetGlobalIndex}(c)$
  \STATE $T[i]\leftarrow$\texttt{False}
  \IF{$\mathtt{IsLeaf}(c)$}
    \FORALL{$j\in\llbracket\lvert\mathcal{C}\lvert\llbracket$}
      \STATE $S_j[c]\leftarrow (\texttt{GetAttributeValue}(c)>\mathcal{C}[j])$
    \ENDFOR
  \ELSE
    \FORALL{$j\in\mathcal\llbracket{f^d}\llbracket$}
      \STATE $c'\leftarrow\mathtt{GetChild}(c,j)$
      \STATE $T[i] \overset{\lor}{\leftarrow}\mathtt{RecursivelyPreProcessTree}(c')$
      \IF{$\neg T[i]$}
          \STATE $k\leftarrow\mathtt{GetGlobalIndex}(c')$
	  \FORALL{$l\in\llbracket\lvert\mathcal{C}\lvert\llbracket$}
          \IF{$\neg{j}$}
            \STATE $S_l[i]\leftarrow S_l[k]$
          \ELSE
            \IF{$\neg(S_l[i] \oplus S_l[k])$}
              \STATE $T[i]\leftarrow\texttt{True}$
            \ENDIF
          \ENDIF
        \ENDFOR
      \ENDIF
    \ENDFOR
  \ENDIF
  \RETURN $T[i]$ 
\end{algorithmic}
\end{algorithm}
Algorithm~\ref{alg:recursively-pre-process} summarizes the
pre-processing stage, for every
cursor position $c$ inside the input hypertree grid $\mathcal{H}$.
The goal of this function is two-fold: first, store the position of
the attribute value of $c$ relative to each of the iso-values in each
of the $S_i$ arrays; second, store the truth value at $T[c]$ to indicate
whether $c$ is intercepted by at least one iso-contour.
When $c$ is coarse, $T[c]$ can be
\texttt{True} only when $c$ has in its descent at least two leaf
cells with opposed signs;
in this case, the $S_i[c]$ values are irrelevant.
In contrast, when $c$ is coarse and $T[c]$ is \texttt{False}, then its
entire descent has the same sign, defining the value stored
in~$S_i[c]$; in this case, $T[c]$ as well as the $S_i[c]$ values
are relevant.
When $c$ is a leaf, $T[c]$ is not meaningful and is assigned
\texttt{False} by default;
in this case, the $S_i[c]$ values are relevant.
As required, Algorithm~\ref{alg:recursively-pre-process} needs neither
geometric nor topological information, hereby allowing for the use of
the lightweight \texttt{TreeGridCursor} for the pre-processing stage.

\begin{algorithm}
\caption{$\mathtt{RecursivelyProcessTree}(s)$}
\label{alg:recursively-process}
\begin{algorithmic}[1]
  \IF{$\mathtt{IsLeaf}(s)$}
    \IF{$\neg \mathtt{Masked}(s)$}
      \FORALL{$i\in\llbracket2^d\llbracket$}
        \IF{$\mathtt{IsOwner}(s,i)$}
	  \STATE $D \leftarrow \mathtt{GenerateDualCell}(s,i)$
          \FORALL{$j\in\llbracket\lvert\mathcal{C}\lvert\llbracket$}
            \STATE $\mathcal{I}\, \overset{+}{\leftarrow} \mathtt{MarchingCube}(D,j)$
          \ENDFOR
        \ENDIF
      \ENDFOR
    \ENDIF
  \ELSE
    \STATE $i\leftarrow\mathtt{GetGlobalIndex}(s)$
    \FORALL{$j\in\llbracket\lvert\mathcal{C}\lvert\llbracket$}
      \FORALL{$k\in\mathcal\llbracket{2^d}\llbracket$}
        \STATE $l\leftarrow\mathtt{GetNeighborGlobalIndex}(s,k)$
        \IF{$T[i] \lor T[l] \lor S_j[i] \neq S_j[l]$}
          \FORALL{$k\in\mathcal\llbracket{f^d}\llbracket$}
            \STATE $\mathtt{RecursivelyProcessTree}(\mathtt{GetChild}(s,k))$
          \ENDFOR
	\RETURN
        \ENDIF
      \ENDFOR
    \ENDFOR
  \ENDIF
\end{algorithmic}
\end{algorithm}

Subsequently, the contouring stage executes the function
\texttt{RecursivelyProcessTree()}, described in
Algorithm~\ref{alg:recursively-process}, upon every cell of
$\mathcal{H}$, using a Moore supercursor~$s$.
For each generated dual cell $D$ and each contour value
$\mathcal{C}[j]$, the call to
$\mathtt{MarchingCube}(D,\mathcal{C}[j])$ returns set of polygons
(possibly empty) that is appended to the iso-contour mesh
$\mathcal{I}$.

\begin{figure}[h]
\centering
\includegraphics[width=0.85\columnwidth]{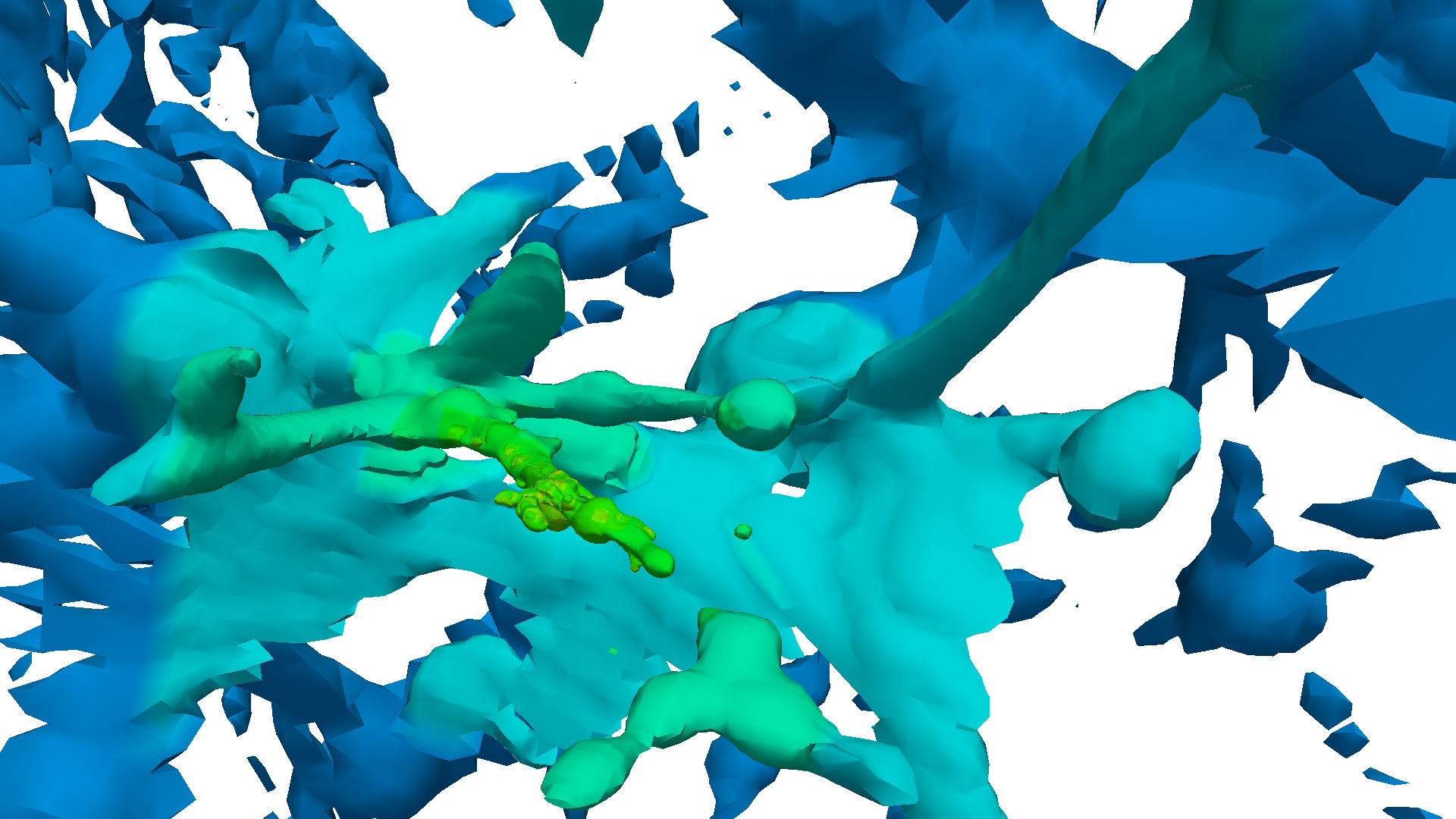}
\caption{Close-up view of an iso-surface generated by the native
\texttt{Contour} filter with a large hypertree grid input.}
\label{fig:IsoAMR}
\end{figure}
Figure~\ref{fig:IsoAMR} illustrates the results of the main
iso-contouring phase, following the pre-processing stage, in the case
of a large AMR simulation.

\section{Results}
\label{s:results}
We now discuss the main results obtained with the
hypertree grid object, beginning with a study of its
performance in terms of memory footprint.
We continue with an overview of the filters that we have developed so
far for this object.
This section ends with a detailed analysis of the axis-aligned
reflection filter, demonstrating the massive memory savings allowed
for by our approach based on separating geometry from topology.
These results are those obtained with our concrete
implementation in \VTK{} version~7.
\subsection{Hypertree Grid Object}
We begin with the case of a hypertree grid with $150$ constituting
hypertrees, used to represent a variable number of cells in an AMR mesh.
\begin{figure}[h]
\centering
\includegraphics[width=0.9\columnwidth]{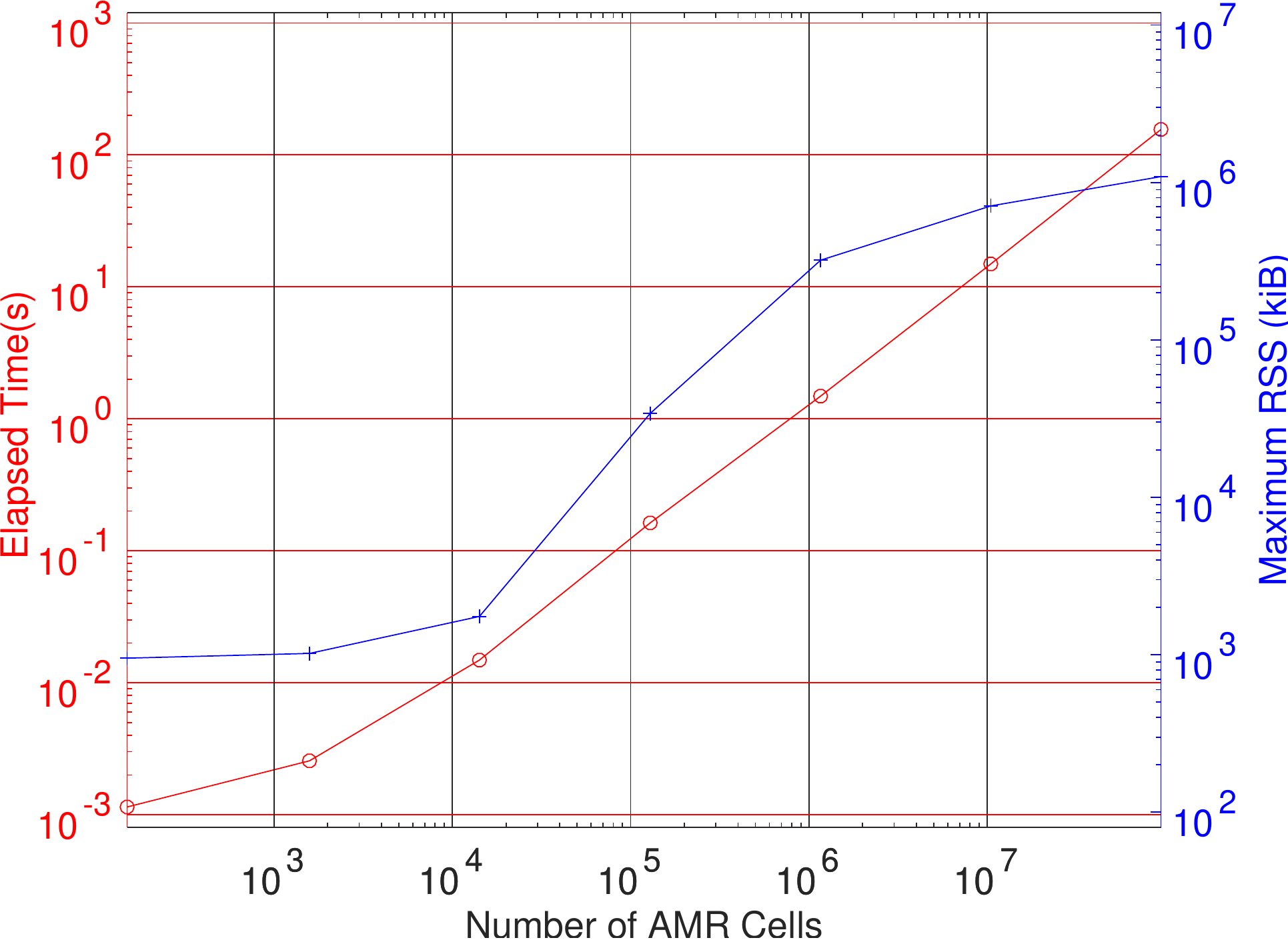}
\caption{Execution time (red) and memory footprint
(blue) \emph{versus} number of cells in a synthetic hypertree grid.}
\label{fig:octant-performance}
\end{figure}

Figure~\ref{fig:octant-performance} illustrates this case, when a varying
number of cells is obtained by increasing the tree depth $\delta\in[1;6]$.
When $\delta=1$, only root-level cells are present in the constituting
hypertrees, resulting in relatively high memory fixed costs per
hypertree; as $\delta$ increase, 
these costs are progressively diluted by the ensuing greater number of
hypertree cells.
In addition, we observe a linear speedup in terms of execution time, hereby
demonstrating the scalability of our approach.

\begin{figure}[h]
\centering
\includegraphics[width=0.9\columnwidth]{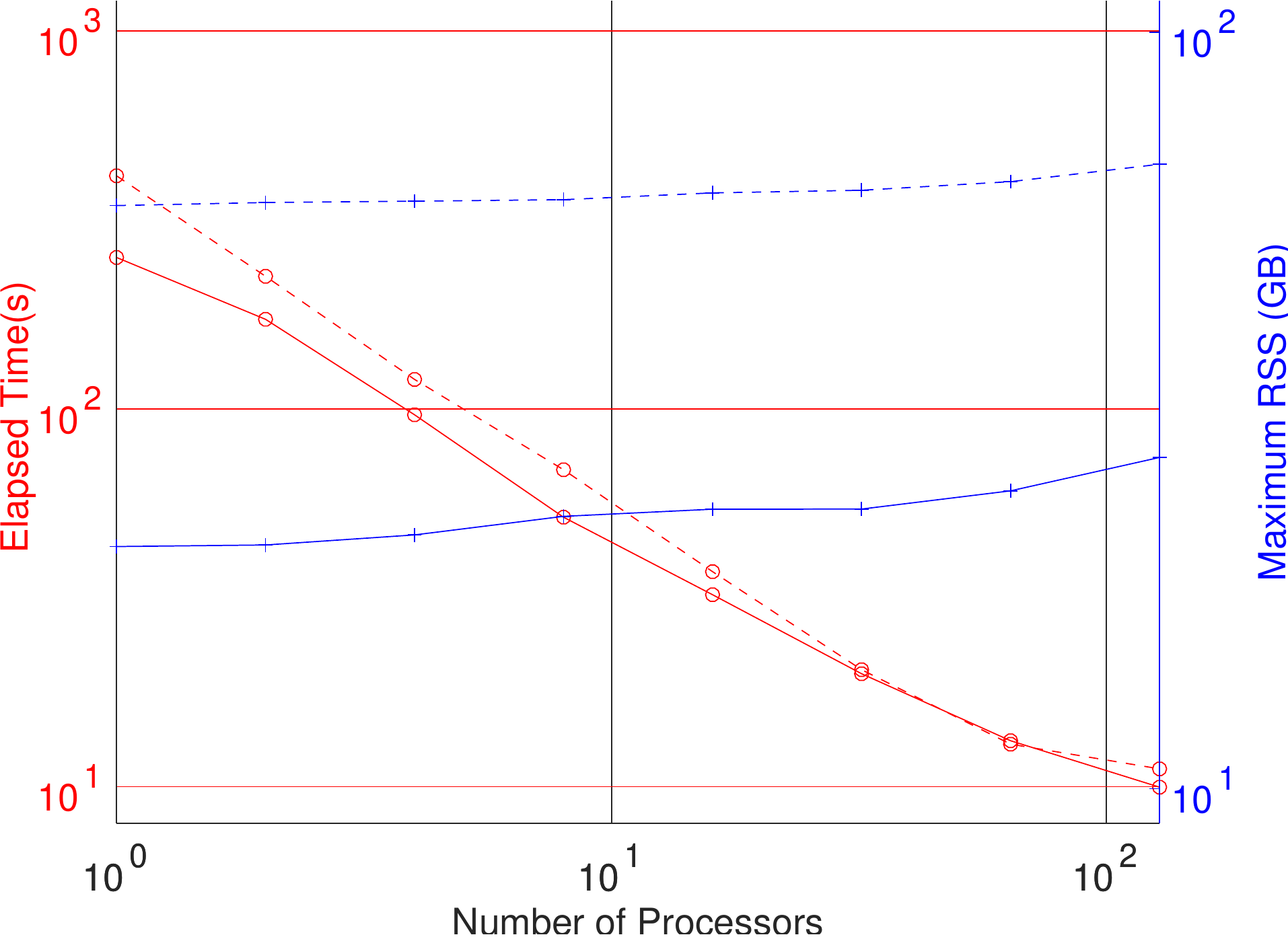}
\caption{Execution time (red) and memory footprint (blue)
\emph{versus} number of processors used to represent a 2D AMR mesh
with $\mathcal{O}(10^8)$ leaves; solid (resp. dashed) lines correspond to
a hypertree (resp. unstructured) grid.}
\label{fig:realAMR-performance}
\end{figure}
Figure~\ref{fig:realAMR-performance} demonstrates the strong
scalability (i.e, with fixed total workload) of our approach, which
scales almost optimally for memory footprint, and super-optimally for
execution time, until maximum speedup is achieved for this problem size.
Moreover, when comparing the performance in terms of memory footprint
of our hypertree grid object with respect to that of using an
unstructured grid representation, we note almost a full order of
magnitude improvement (approximately a factor of $7$).
Furthermore, we have observed that this massive decrease in memory
usage remains constant across a wide range of workload distribution
schemes for highly refined meshes.
 
\subsection{Hypertree Grid Filters}
We now illustrate some results obtained with the native hypertree grid
filters presented in~\S\ref{s:concrete-filters} and implemented
in~\VTK{}, exploring the $2$ and $3$-dimensional cases as well as
the two possible branch factor values.
\begin{figure}[htb!]
\centering
\begin{minipage}[t]{0.32\columnwidth}
\centering
\vspace{0pt}
\includegraphics[width=0.99\columnwidth]{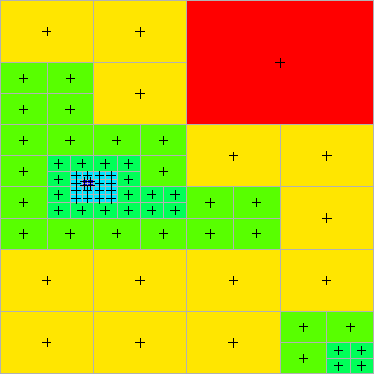}\\
(a)\\[2mm]
\includegraphics[width=0.99\columnwidth]{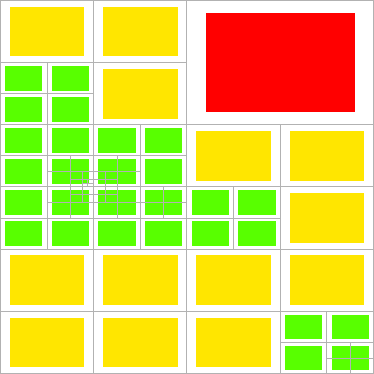}\\
(d)\\[2mm]
\includegraphics[width=0.99\columnwidth]{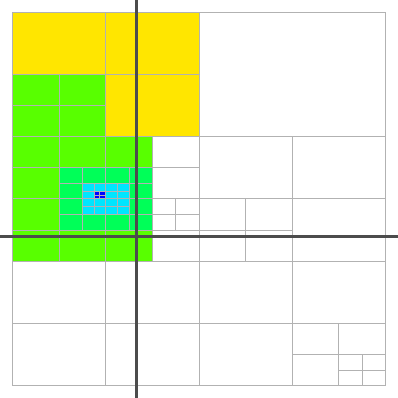}\\
(g)
\end{minipage}
\hfil
\begin{minipage}[t]{0.32\columnwidth}
\centering
\vspace{0pt}
\includegraphics[width=0.99\columnwidth]{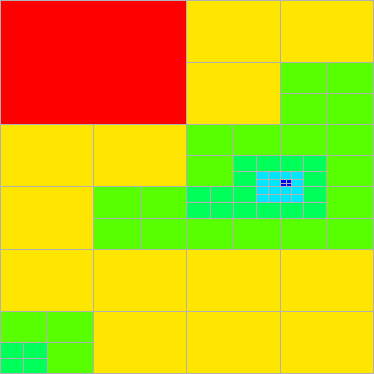}\\
(b)\\[2mm]
\includegraphics[width=0.99\columnwidth]{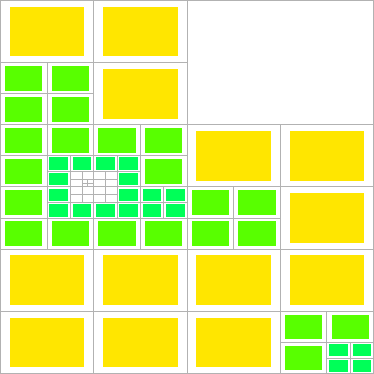}\\
(e)\\[2mm]
\includegraphics[width=0.99\columnwidth]{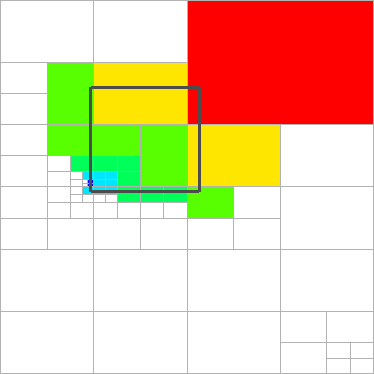}\\
(h)
\end{minipage}
\hfil
\begin{minipage}[t]{0.32\columnwidth}
\centering
\vspace{0pt}
\includegraphics[width=0.99\columnwidth]{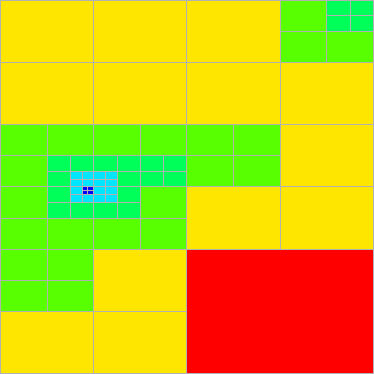}\\
(c)\\[2mm]
\includegraphics[width=0.99\columnwidth]{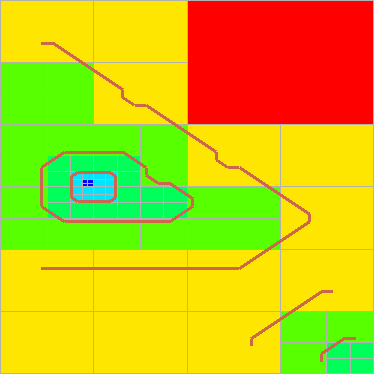}\\
(f)\\[2mm]
\includegraphics[width=0.99\columnwidth]{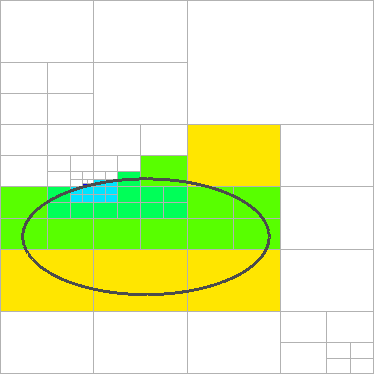}\\
(i)
\end{minipage}
\caption{Visualizations of the data sets produced by application of
native filters to a $2$-dimensional, binary hypertree grid with 6 root
cells. 
In all of these, the \texttt{Geometry} filter was used in order to
display the underlying AMR mesh, where colors represent cell depth
used a cell-centered attribute:
\texttt{CellCenters} (a),
\texttt{AxisReflection} (b\&c), respectively across vertical and horizontal center lines 
\texttt{DepthLimiter} (d)
\texttt{Threshold} (e),
\texttt{Contour} (f),
\texttt{AxisClip} (g) through (i), respectively by 2 lines parallel to
the grid axes, a axis-aligned rectangle, and an ellipse.}
\label{fig:HyperTreeGridBinary2DFilters}
\end{figure}

We begin with visualizations obtained when the input data set
is a 2-dimensional binary hypertree grid,
with a $2\times3$ layout of root cells, to which is attached a single
attribute field filled with the cell depths.
Figure~\ref{fig:HyperTreeGridBinary2DFilters} applies the native hypertree
grid \texttt{Geometry} filter (note that shrinkage of the output
geometry is sometimes used in order to facilitate the interpretation
of the results) to render hypertree grid outputs.
In addition to it, these images illustrate the following filters:
\begin{description}
\item[(a)]
\texttt{CellCenters}, hooked to a glyphing filter to
produce the black crosses shown at cell centers. 
\item[(b\&c)]
\texttt{AxisReflection}, where  hyperplanes are lines, respectively
parallel to the vertical and horizontal axes, passing through the
center of the hypertree grid.
\item[(d)]
\texttt{DepthLimiter} with depth limit is set to~$2$.
\item[(e)]
\texttt{Threshold} for attribute values within $[1;3]$.
\item[(f)]
\texttt{Contour} with attribute iso-values $1.25$, $2.5$,
and $3.75$.
Note that, as explained in~\S\ref{s:dual}, the contours are
topologically correct but do not intercept the primal boundary,
because a non-adjusted dual is used.
\item[(g--i)]
\texttt{AxisClip}, illustrated for each of its
three modes of operation, respectively: hyperplane (here, with 2
consecutive appelications), box, and a quadratic corresponding
to an axis-aligned ellipse; note that alignment
with the grid axes is not required by the filter as any arbitrary
quadratic can be specified.
\end{description}

\begin{figure}[htb!]
\centering
\begin{minipage}[t]{0.32\columnwidth}
\centering
\vspace{0pt}
\includegraphics[width=0.99\columnwidth]{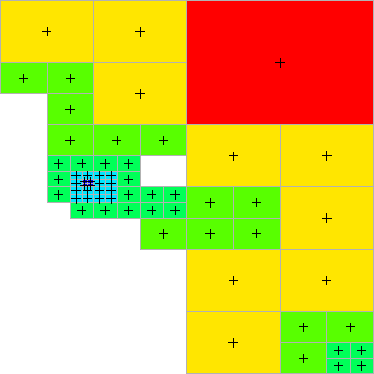}\\
(a)\\[2mm]
\includegraphics[width=0.99\columnwidth]{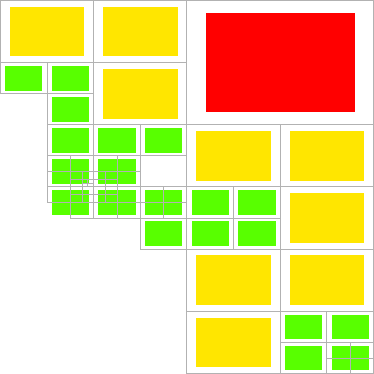}\\
(d)
\end{minipage}
\hfil
\begin{minipage}[t]{0.32\columnwidth}
\centering
\vspace{0pt}
\includegraphics[width=0.99\columnwidth]{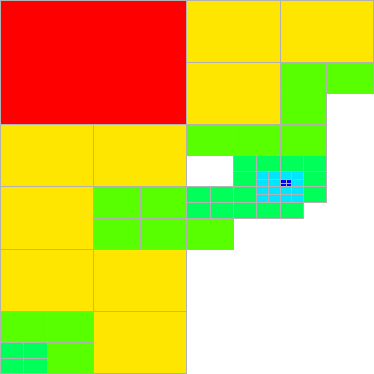}\\
(b)\\[2mm]
\includegraphics[width=0.99\columnwidth]{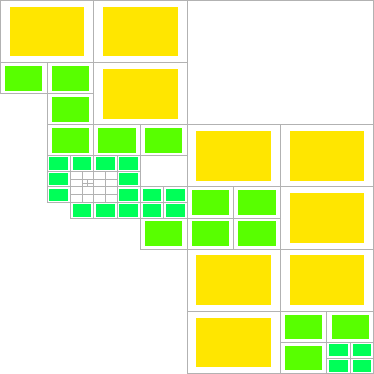}\\
(e)
\end{minipage}
\hfil
\begin{minipage}[t]{0.32\columnwidth}
\centering
\vspace{0pt}
\includegraphics[width=0.99\columnwidth]{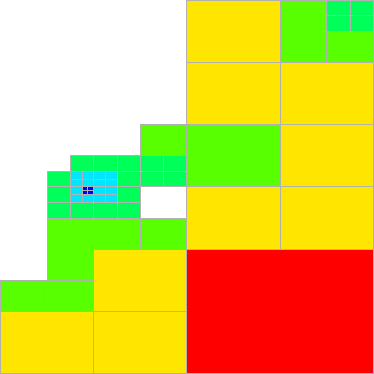}\\
(c)\\[2mm]
\includegraphics[width=0.99\columnwidth]{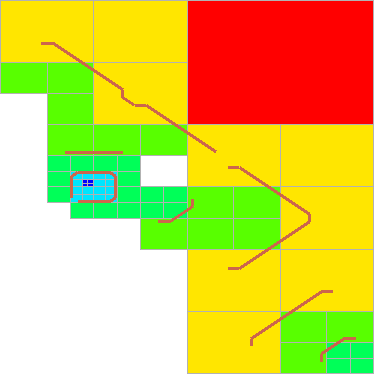}\\
(f)
\end{minipage}
\caption{Results of the same tests as the first six shown
Figure~\ref{fig:HyperTreeGridBinary2DFilters}, but when a non-empty
material mask was added to the input grid.}
\label{fig:HyperTreeGridBinary2DMaterialFilters}
\end{figure}
The results computed by the same filters, but when a
non-empty material mask is attached to the hypertree grid input, are
shown in Figure~\ref{fig:HyperTreeGridBinary2DMaterialFilters}.
We are not showing here the results obtained with the
\texttt{AxisClip} filters in order to save space, but suffices to say
that corresponding images are obtained are expected.

Of particular interest is the iso-contouring case (f): because the
current implementation of the filter uses a non-adjusted dual, the
computed iso-contours exhibit additional geometric oddities in the
vicinity of the non-convexities resulting from the presence of masked
cells.
In our typical, large-scale applications, the phenomena of interest which
are searched for in the post-processing stage tend to be removed from
object boundaries (external or internal); therefore, geometric
error in computed iso-contours are generally not encountered.
It however remains our goal to provide the option to adjust the dual
in future implementations.

The case of the \texttt{DepthLimiter} filter (d) also reveals an
interesting feature, that can only present itself when non-convexities
are present -- and therefore, only when a non-empty mask is attached
to the input hypertree grid.
Specifically, the hypertree grid output by the filter can have a
larger geometric extent that the input.
This results from the fact that an input coarse cell at the depth
limit is retained to create an output leaf as soon as \emph{at least}
one of its descendents is not masked.
Indeed, this behavior can be observed in the figure, with the green
cell at depth~$2$ located at the middle-left of the grid.

\begin{figure}[htb!]
\centering
\begin{minipage}[t]{0.32\columnwidth}
\centering
\vspace{0pt}
\includegraphics[width=0.85\columnwidth]{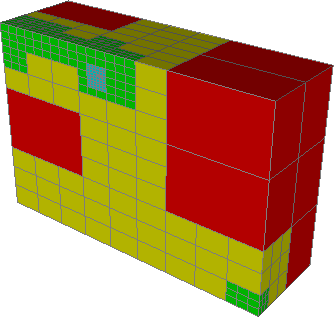}\\
(a)\\[2mm]
\includegraphics[width=0.85\columnwidth]{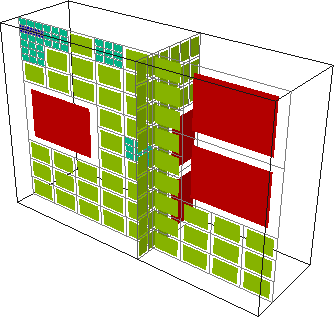}\\
(d)\\[2mm]
\includegraphics[width=0.85\columnwidth]{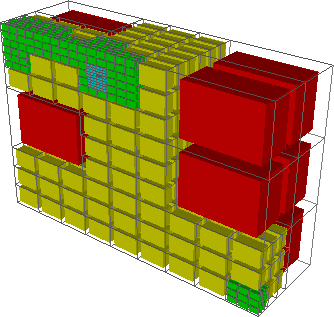}\\
(g)\\[2mm]
\includegraphics[width=0.85\columnwidth]{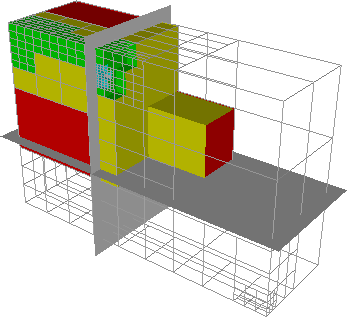}\\
(j)
\end{minipage}
\hfil
\begin{minipage}[t]{0.32\columnwidth}
\centering
\vspace{0pt}
\includegraphics[width=0.85\columnwidth]{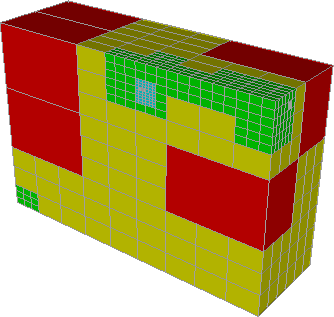}\\
(b)\\[2mm]
\includegraphics[width=0.85\columnwidth]{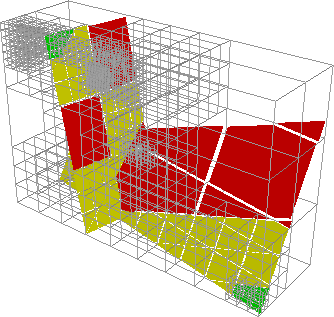}\\
(e)\\[2mm]
\includegraphics[width=0.85\columnwidth]{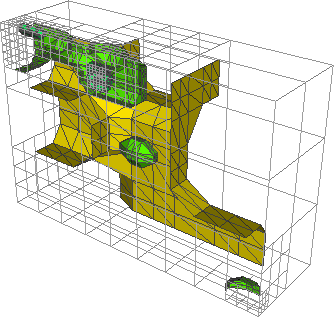}\\
(h)\\[2mm]
\includegraphics[width=0.85\columnwidth]{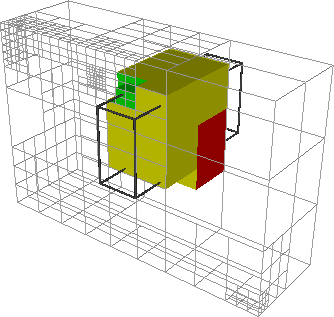}\\
(k)
\end{minipage}
\hfil
\begin{minipage}[t]{0.32\columnwidth}
\centering
\vspace{0pt}
\includegraphics[width=0.85\columnwidth]{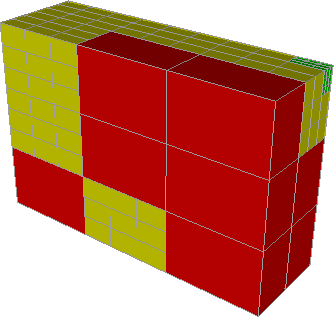}\\
(c)\\[2mm]
\includegraphics[width=0.85\columnwidth]{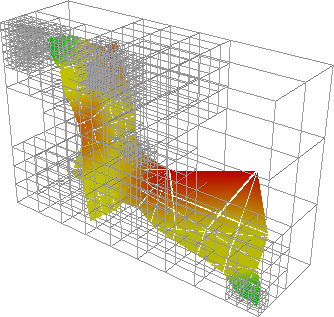}\\
(f)\\[2mm]
\includegraphics[width=0.85\columnwidth]{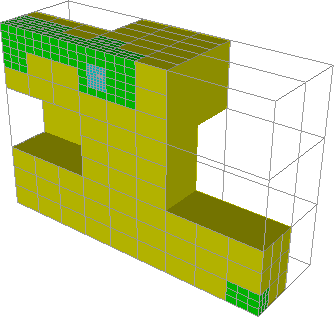}\\
(i)\\[2mm]
\includegraphics[width=0.85\columnwidth]{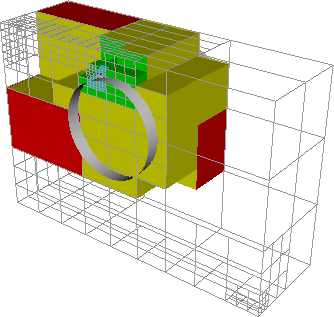}\\
(l)
\end{minipage}
\caption{Renderings of the outputs of hypertree grid filters to a
$3$-dimensional, ternary hypertree grid with 18 root 
cells. 
The hypertree grid \texttt{Geometry} filter was used to represent the
input AMR mesh in wireframe mode wherever visible, as well as the
output hypertree grid objects whenever applicable, where
colors represent cell depth:
\texttt{Geometry} (a),
\texttt{AxisReflection} (b\&c), respectively across one and two
axis-aligned planes, 
\texttt{AxisCut} (d), 
\texttt{PlaneCutter} (e\&f), respectively in primal and dual mode,
\texttt{ToUnstructured} (g), 
\texttt{Contour} (h), 
\texttt{Threshold} (i),
(j--l): \texttt{AxisClip}, respectively by 2 planes parallel to
the grid axes, an axis-aligned box, and a cylinder.}
\label{fig:HyperTreeGridTernary3DFilters}
\end{figure}

A $3$-dimensional, ternary set of test cases is now used,
with a $3\times3\times2$ layout of roots to further illustrate our point.
In Figure~\ref{fig:HyperTreeGridTernary3DFilters}, we show
visualizations obtained with the following filters (note that
\texttt{Geometry} is used to visualize all hypertree grid outputs):
\begin{description}
\item[(a)]
\texttt{Geometry}.
\item[(b\&c)]
\texttt{AxisReflection}, respectively with 1 and 2 successive
reflections about planes passing through the center of the
hypertree grid.
\item[(d)]
\texttt{AxisCut} with two axis-aligned cut planes, producing two
$2$-dimensional hypertree grids, whose geometry is shrunk for legibility.
\item[(e\&f)]
\texttt{PlaneCutter} which, in contrast, produces polygonal data sets,
respectively in primal and dual modes.
The main benefit of latter is its conforming mesh output, and it
should thus always used when subsequent post-processing requiring
perfect connectivity is intended; it is however important to note that
is not only less visually appealing, but also considerably slower than
the former.
\item[(g)]
\texttt{ToUnstructured}, whose all-hexahedral unstructured grid output
is connected downstream to a shrink filter.
As discussed in~\S\ref{s:problem}, the resulting unstructured mesh is
not conforming, because interior faces are not shared but replicated.
\item[(h)]
\texttt{Contour}, again with three iso-values.
\item[(i)]
\texttt{Threshold}, for depth values within $[1;3]$.
\item[(j--l)]
\texttt{AxisClip} with its three modes of operation:
respectively, two successive clips with planes parallel to the grid
axes, an axis-aligned box, and a quadratic associated with a cylinder
of revolution about the third coordinate axis.
\end{description}

\begin{figure}[h!]
\centering
\begin{minipage}[t]{0.32\columnwidth}
\centering
\vspace{0pt}
\includegraphics[width=0.85\columnwidth]{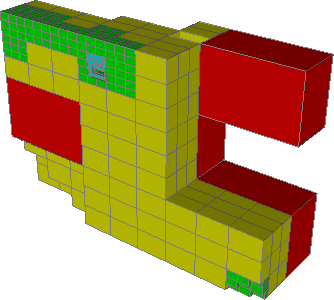}\\
(a)\\[2mm]
\includegraphics[width=0.85\columnwidth]{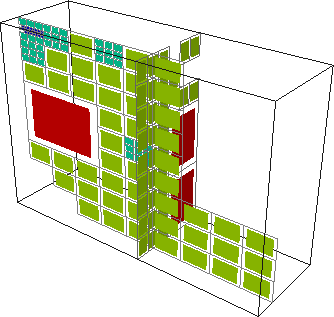}\\
(d)\\[2mm]
\includegraphics[width=0.85\columnwidth]{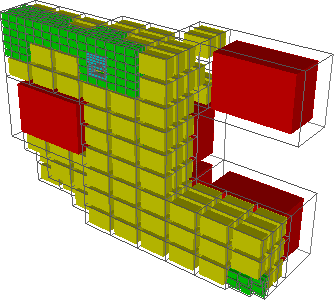}\\
(g)
\end{minipage}
\hfil
\begin{minipage}[t]{0.32\columnwidth}
\centering
\vspace{0pt}
\includegraphics[width=0.85\columnwidth]{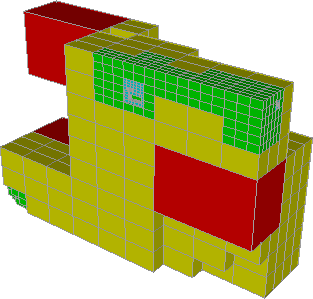}\\
(b)\\[2mm]
\includegraphics[width=0.85\columnwidth]{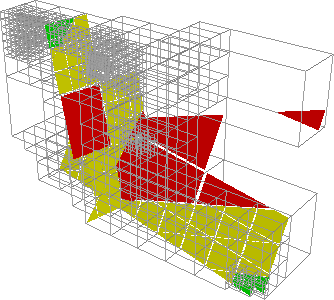}\\
(e)\\[2mm]
\includegraphics[width=0.85\columnwidth]{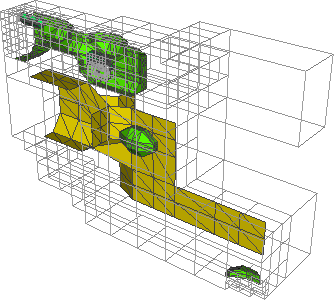}\\
(h)
\end{minipage}
\hfil
\begin{minipage}[t]{0.32\columnwidth}
\centering
\vspace{0pt}
\includegraphics[width=0.85\columnwidth]{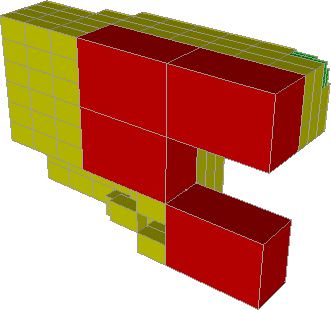}\\
(c)\\[2mm]
\includegraphics[width=0.85\columnwidth]{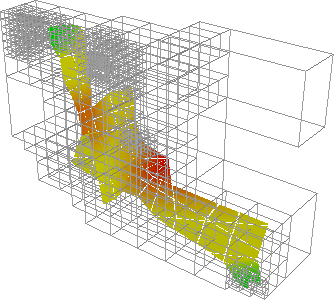}\\
(f)\\[2mm]
\includegraphics[width=0.85\columnwidth]{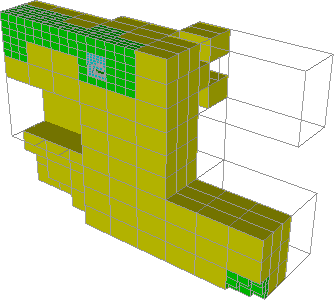}\\
(i)
\end{minipage}
\caption{Results of the same tests as the first six shown
Figure~\ref{fig:HyperTreeGridTernary3DFilters}, but when a non-empty
material mask was added to the input grid.}
\label{fig:HyperTreeGridTernary3DMaterialFilters}
\end{figure}

As previously done with the $2$-dimensional cases,
Figure~\ref{fig:HyperTreeGridTernary3DMaterialFilters} present a
subset of these cases, but obtained with a non-empty mask attached to
the input hypertree grids.
Comments similar to those made in the $2$-dimensional, binary case can
be made and we will not therefore repeat ourselves.
The interested reader is invited to draw parallels between
corresponding $2$ and $3$ dimensional sub-figures, and to inspect the
contents of the test harness we implemented for all existing hypertree
grid filters, across different dimensions, branching factors, and other modalities:
to date, 58 individual tests are available and can be either executed
as they are, or modified and experimented with at will.

\begin{figure}[h]
\centering
\includegraphics[height=35mm]{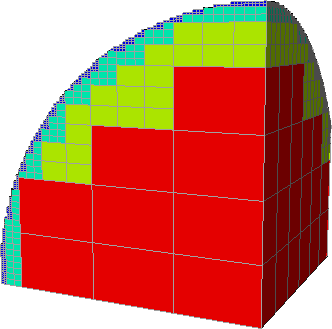}
\hfil
\includegraphics[height=35mm]{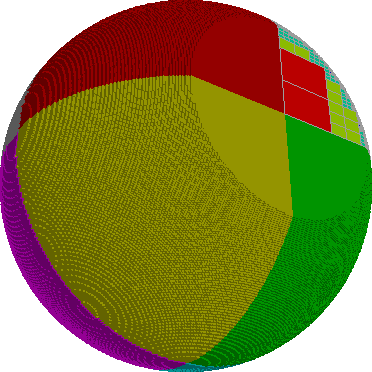}
\caption{Left: the first octant of a truncated unit ball, approximated
with 5 levels of a ternary tree-based AMR grid with $5\times5\times6$ root cells.
Right: a rendering showing the same truncated octant (upper right corner),
together with its successive images (in solid colors) by axis-aligned
reflections, yielding a truncated unit ball.}
\label{fig:SphereReflections}
\end{figure}
We close this discussion with the particular case of the
\texttt{AxisReflection} filter.
In Figure~\ref{fig:SphereReflections}, we illustrate the use of this
filter by applying it to the case of a ternary tree-based AMR grid
with $5\times5\times6$ root cells, where a material mask is defined using a
quadratic function retaining only those cells that are within
or intersect a truncated octant of the unit ball.
The experiment thus consisted in creating this object, hereafter
referred to as the \texttt{octant}, then performing seven
reflection across planes, adequately defined in order to produce
outputs whose union, together with the initial
\texttt{octant}, produces the unit ball truncated by the original
plane and its symmetrical about the sphere center.
This output is referred to as the \texttt{reflections}.
Octant creation, geometry extraction and rendering times were excluded
from this experiment in order to assess the performance of the
\texttt{AxisReflection} filter in isolation.

\begin{table}[htb]
\renewcommand{\arraystretch}{1.2}
\caption{Main characteristics, and memory footprints in terms of
maximum resident set size, of a ternary hypertree grid object
(\texttt{octant}), its $8$-time replication (\texttt{octant*8}) and of
its union (\texttt{reflections}) with 7 images thereof by the
\texttt{AxisReflection} filter.}
\label{t:axis-reflection-results}
\centering
  \begin{tabular}{@{}lrrrr@{}}
  \hline \\[-12pt]
  &Number   &Number    &Number   &\textbf{RSS} \\
  &of cells &of leaves &of trees &\textbf{(kiB)}   \\
  \hline\hline
  \texttt{octant} 
  &  $128724$ &$123962$ & $150$ &$\textbf{ 44924}$ \\
  \hline
  \texttt{octant*8} 
  & $1029792$ &$991696$ &$1200$ &$\textbf{359392}$ \\
  \hline
  \texttt{reflections}
   &$1029792$ &$991696$ &$1200$ &$\textbf{ 45172}$ \\
  \hline
  \end{tabular}
\end{table}
The main results of this experiment performed on a single core are
summarized in Table~\ref{t:axis-reflection-results}; in particular,
the \texttt{reflections} represents an AMR mesh 8 times larger, with over
one million cells (96.3\% of which are leaf cells), than the original
\texttt{octant}.
Executing the 7 reflections took a negligible time, compared to the
octant creation or its rendering, hereby confirming the theoretical
prediction that, if correctly implemented, the reflection filter
should have negligible execution time.
Another key finding of this test was to measure a negligible increase
in memory readings\footnote{We assess memory footprint in terms of
maximum resident set size (RSS), indicating the amount of memory that
belongs to a process and resides in RAM.}, as compared to the real
replication of the object requiring a commensurate increase in memory
footprint (which might not be available to the target platform).
These results demonstrate that our implementation fully delivers the
promises of the theoretical analysis, in terms of execution speed as
well as of memory footprint.
As a result, all future hypertree grid structure-preserving geometry
transformation filters shall be implemented following the same paradigm.

\section{Conclusion}
\label{s:conclusion}
There are many more details to this story than we could
possibly fit within the frame of a journal article.
What are we, then, to make of this already long \emph{expos\'e},
which encompassed general motivations, theoretical foundations,
application methods, and experimental results?

In the next few lines, we will first look back on what has been
achieved so far, as compared to what we were initially envisioning.
This will allow us to conclude with a set of remarks as to
our subsequent projects and goals, articulating them within our
general vision together with what we have discovered and done during
the course of the work described in this article.
\subsection{Main Findings}
We set out in~\S\ref{s:vision} ([a]) with the goal to propose a novel
\VTK{} data object that would be able to support all conceivable types of
rectilinear, tree-based AMR data sets, based not only on today's
software but also on what we can foresee of tomorrow's extreme-scale
simulations.
We can confidently claim that we have accomplished this first goal,
based on the \texttt{vtkHyperTreeGrid} object and its family of
lesser objects presented throughout this article.
In particular, the key design constraint to drastically reduce memory
usage, as compared to either earlier implementations of this object or
to different, existing \VTK{} data objects, was fully achieved.
This was amply demonstrated by our numerical results in
in~\S\ref{s:results}, consistent with what the theory laid out in
~\S\ref{s:foundations} was allowing to hope for.
This achievevement dramatically reduces hardware requirements by
several orders of magnitude, as compared to the
alternatives currently used in AMR visualization.

Moreover, we propounded in \S\ref{s:vision} ([b]) to design
and implement visualization filters that could natively
operate on this novel data object, with the added stated goal of
measurable performance in terms of execution speed.
We have also entirely fulfilled our objectives in this regard, with
demonstrated performance improvements with respect to our earlier
design and implementation (not to mention the alternatives which are
plainly useless for large-scale meshes). 

Meanwhile, in the process of designing such filters, we entirely
revised our earlier notion of tree cursors and, more importantly,
supercursors.
This resulted, in particular, in the introduction
of a hierarchy of such objects in order
to allow for the selection of the cursor that is 
the most tightly adapted to the particular algorithm being considered.
Due in particular to use of templates, as well to the careful nesting
of geometric and topological cursor properties, this results in the
added benefits of enhanced code maintainability and legibility.
This new incarnation of the hypertree grid object allows us to
confidently assert that it can be used even by an application
developer not intimately familiar with the implementation details.

An other important aspect of this work is the availability of a full
set of visualization filters able to natively operate over the novel
data object.
As explained this article, these filters have all been written with
performance in mind, in terms of both memory footprint and execution
speed.
Furthermore, our design based on the hierarchy of templated cursors and
supercursors, combined with the pre-selection paradigm for enhanced
performance, gives developers the opportunity to easily create new
filters tailored to their particular needs.

After all has been said in terms of theoretical soundness, or of
convincing experimental results, what is ultimately our main claim
to success is that our HERA code users at CEA have been able to
begin routine post-processing of their AMR simulations, within a
setting similar to that which already exists for the visualization of
simulations based on other types of meshes.

Based on these findings, we decided
to contribute our development to the \VTK{} code base so it can
benefit the tree-based AMR community at large.

\subsection{Perspectives}
As formulated in~\S\ref{s:vision}, we have considered many potential
avenues for further advances in the field of tree-based AMR
visualization and analysis.
These are not only of academic interest; in fact, they appear as
strictly necessary when considering the post-processing options that
are commonly available for other types of simulations, such as the
finite element method using fully unstructured, conforming meshes.

First, we expecte that our original goal [b] will be further achieved,
as the community of users of our contributed code will increase and
expand to connected yet different application domains.

Meanwhile, 2-dimensional AMR visualization can be especially
challenging, as it requires that all leaf cells be rendered.
In consequence, the interactivity of the visualization process
decreases as input data object size increases.
This problem is further compounded by the enhanced efficiency, in terms of
memory footprint, of our hypertree grid model which elicits a new
situation where rendering has become the bottleneck for our the
target platforms.
As a result, the next goal ([c]) is indeed an urgent need, for which
the lack of existing solution is currently hindering the AMR
visualization and analysis workflow.
Our preliminary developments in this regard should be finalized,
validated and contributed shortly.
These focus on rendering speed, in particular in dimension~2, by
exploiting level-of-detail properties, which we also plan to carefully
study and explain in a sequel to this article.

Besides, the $3$-dimensional visualization technique known as volume
rendering, which has now been broadly used for almost two decades, for
different types of data objects, remains mostly unchartered territory
when it comes to tree-based AMR data and would come in direct support
of our stated goal ([d]).
Iso-contouring is often derided as being the ``poor man's volume
rendering''.
Albeit excessive, as in many cases an iso-surface is exactly what is
required by the nature of the analysis being performed, this statement
nonetheless usefully conveys the general idea that ``true''
3-dimensional visualization is a capability that most if not all
users want to have in a visualization tool set before they deem it
sufficient.
Considerable theoretical and experimental effort will be required in 
order to support this need, for almost no prior work exists in this area.
However, such a major endeavor could potentially be amortized by 
2-dimension specializations in addition to the overarching
3-dimensional goal. 

The work done so far does not address \emph{per se} any of items
[e]-[g] in our initial vision.
However, we believe that the theoretical groundwork which we have
already conducted will allow for an easier pursuit of these goals in
the future.

Last, we would like to close this panorama by mentioning an ongoing
reflection regarding \emph{in situ} and \emph{in transit}
visualization and analysis.
This contemplates the possibilities that exist
to directly couple an existing production-level AMR simulation code
with a visualization tool set adapted to it, in a fashion that would
entirely eliminate intermediate storage to disk.
We are confident that this will allow us to address the
last vision item ([h]) in the near future, which will be
discussed in subsequent work.
		
\subsection*{Acknowledgments}
This work was supported by the French Alternative Energies and Energy Atomic Commission (CEA)
for Military Applications (DAM).
We are grateful for advice from C. Guilbaud and N. Lardjane (CEA)
and helpful discussions with J.-C. Frament (Positiveyes).
We also thank D. Chapon at the Institute of Research into the
Fundamental Laws of the Universe (Irfu) for providing RAMSES
large-scale simulation results and helping us exploit these to
generate the astrophysics-related visualization results shown in this
article.
This article could not have been written without the support and
patience of ours families. 
\bibliographystyle{plain}
\bibliography{ewc17}

\begin{thebibliography}{10}

\bibitem{tera:CEA}
{CEA}'s {Tera} supercomputer.
\newblock http://www-hpc.cea.fr/en/complexe/tera.htm.

\bibitem{chombo:overview}
Overview of block structured {AMR}.
\newblock Online: https://commons.lbl.gov/display/chombo/
  Overview+of+Block+Structured+AMR.

\bibitem{lomov:05}
Patch-based adaptive mesh refinement for multimaterial hydrodynamics.
\newblock In {\em Joint Russian-American Five-Laboratory Conference on
  Computational Mathematics/Physics}, Vienna, Austria, June 2005.

\bibitem{aguilera:07}
D.~Aguilera, T.~Carrard, G.~Colin de~Verdi\`ere, J.-P. Nomin\'e, and
  V.~Tabourin.
\newblock Parallel software and hardware for capability visualization of {HPC}
  results.
\newblock {\em Numerical Modeling of Space Plasma Flows: Astronum}, 2007.

\bibitem{love:12}
D.~Aguilera, T.~Carrard, C.~Guilbaud, J.~Schneider, and S.~Sorbet.
\newblock Visualization and post-processing for high performance computing.
\newblock {\em CHOCS}, 41:57--67, 2012.

\bibitem{avila:10}
L.~Avila, U.~Ayachit, S.~Barr\'e, J.~Baumes, F.~Bertel, R.~Blue, D.~Cole,
  D.~DeMarle, B.~Geveci, W.~Hoffman, B.~King, K.~Krishnan, C.~Law, K.~Martin,
  W.~McLendon, P.~P\'ebay, N.~Russell, W.~Schroeder, T~Shead, J.~Shepherd,
  A.~Wilson, and B.~Wylie.
\newblock {\em The {VTK} User's Guide}.
\newblock Kitware, Inc., eleventh edition, 2010.

\bibitem{bale:02}
D.~Bale, R.~J. LeVeque, S.~Mitran, and J.~A. Rossmanith.
\newblock A wave-propagation method for conservation laws and balance laws with
  spatially varying flux functions.
\newblock {\em SIAM J. Sci. Comput.}, 24:955--978, 2002.

\bibitem{berger:11}
M.~J. Berger, D.~L. George, R.~J. LeVeque, and K.~T. Mandli.
\newblock The {GeoClaw} software for depth-averaged flows with adaptive
  refinement.
\newblock {\em Adv. Water Res.}, 34:1195--1206, 2011.

\bibitem{berger:84}
M.~J. Berger and J.~Oliger.
\newblock Adaptive mesh refinement for hyperbolic partial differential
  equations.
\newblock {\em J. Comput. Phys.}, 53(3):484--512, 1984.

\bibitem{bondy:11}
A.~Bondy and U.S.R. Murty.
\newblock {\em Graph Theory}.
\newblock Graduate Texts in Mathematics. Springer London, 2011.

\bibitem{hercule:12}
O.~Bressand, L.~Colombet, A.~Fontaine, G.~Harel, and J.-B. Lekien.
\newblock Hercule: A library of scientific data management for numerical
  simulation.
\newblock {\em CHOCS}, 41:29--37, 2012.

\bibitem{carrard:imr21}
T.~Carrard, C.~Law, and P.~P\'eba\"y.
\newblock A generic hyper tree grid implementation for {AMR} mesh manipulation
  and visualization in {VTK}.
\newblock In {\em Proc. 21$^{st}$ International Meshing Roundtable}, San Jose,
  CA, U.S.A., October 2012.

\bibitem{delaunay:34}
B.~Delaunay.
\newblock Sur la sph\`ere vide.
\newblock {\em Bul. Acad. Sci. URSS, Class. Sci. Nat.}, pages 793--800, 1934.

\bibitem{frey:08}
P.~Frey and P.-L. George.
\newblock {\em Mesh generation}.
\newblock John Wiley \& Sons, 2 edition, 2008.

\bibitem{flash:00}
B.~Fryxell, K.~Olson, P.~Ricker, F.~X. Timmes, M.~Zingale, D.~Q. Lamb,
  P.~MacNeice, R.~Rosner, J.~W. Truran, and H.~Tufo.
\newblock {FLASH}: An adaptive mesh hydrodynamics code for modeling
  astrophysical thermonuclear flashes.
\newblock {\em The Astrophysical Journal Supplement Series}, 131(1):273, 2000.

\bibitem{gittings:08}
M.~Gittings, R.~Weaver, M.~Clover, T.~Betlach, N.~Byrne, R.~Coker, E.~Dendy,
  R.~Hueckstaedt, K.~New, W.~R. Oakes, D.~Ranta, and R.~Stefan.
\newblock The {RAGE} radiation-hydrodynamic code.
\newblock {\em Computational Science \& Discovery}, 1(1), 2008.

\bibitem{jourdren:05}
H.~Jourdren.
\newblock {HERA}: A hydrodynamic {AMR} platform for multi-physics simulations.
\newblock In {\em Adaptive Mesh Refinement Theory and Application}, volume~41
  of {\em LNCSE}, pages 283--294. Springer, 2005.

\bibitem{packard:85}
N.~Packard and S.~Wolfram.
\newblock Two dimensional cell automata.
\newblock {\em J. Comput. Phys.}, (38):901--946, 1985.

\bibitem{ramses:02}
R.~Teyssier.
\newblock Cosmological hydrodynamics with adaptive mesh refinement. a new high
  resolution code called {RAMSES}.
\newblock {\em Astronomy and Astrophysics}, 385:337--364, 2002.

\bibitem{woodward:12}
P.~Woodward, J.~Jayayaraj, P.H. Lin, Mike M.~Knox, D.~Porter, C.~Fryer,
  G.~Dimonte, C.~Joggerst, G.~Rockefeller, W.~Dai, R.~Kares, and V.~Thomas.
\newblock Simulating turbulent mixing from {Richtmyer-Meshkov and
  Rayleigh-Taylor} instabilities in converging geometries using moving
  cartesian grids.
\newblock Technical Report LA-UR-13-20949, Los Alamos National Laboratory,
  2012.

\bibitem{yau:83}
M.-M. Yau and S.~N. Srihari.
\newblock A hierarchical data structure for multidimensional digital images.
\newblock {\em Communications of the ACM}, 26(7):504--515, July 1983.

\end{thebibliography}
\end{document}